\documentclass[sigconf]{acmart}

\usepackage{textcomp}
\usepackage{stfloats}
\usepackage{url}
\usepackage{verbatim}
\usepackage{amsthm}
\usepackage{subfigure}
\usepackage{makecell}
\usepackage{multirow}
\usepackage{enumitem}
\usepackage{subcaption}

\usepackage{color, colortbl}

\definecolor{Gray}{rgb}{0.9, 0.9, 0.9}

\newcommand{\gray}{\cellcolor{Gray}}

\usepackage[capitalize,noabbrev]{cleveref}

\usepackage[ruled,linesnumbered]{algorithm2e}

\usepackage[inkscapelatex=false]{svg}

\AtBeginDocument{%
  }

\copyrightyear{2025}
\acmYear{2025}
\setcopyright{acmlicensed}\acmConference[KDD '25]{Proceedings of the 31st ACM SIGKDD Conference on Knowledge Discovery and Data Mining V.2}{August 3--7, 2025}{Toronto, ON, Canada}
\acmBooktitle{Proceedings of the 31st ACM SIGKDD Conference on Knowledge Discovery and Data Mining V.2 (KDD '25), August 3--7, 2025, Toronto, ON, Canada}
\acmDOI{10.1145/3711896.3737042}
\acmISBN{979-8-4007-1454-2/2025/08}

\begin{document}

\title[MDVT]{MDVT: Enhancing Multimodal Recommendation with Model-Agnostic Multimodal-Driven Virtual Triplets}

\author{Jinfeng Xu}
\email{jinfeng@connect.hku.hk}
\affiliation{%
  \institution{Department of Electrical and Electronic Engineering, \\The University of Hong Kong}
  \city{HongKong SAR}
  \country{China}}

\author{Zheyu Chen}
\email{zheyu.chen@connect.polyu.hk}
\affiliation{%
  \institution{Department of Electrical and Electronic Engineering, \\The Hong Kong Polytechnic University}
  \city{HongKong SAR}
  \country{China}}

\author{Jinze Li}
\email{lijinze-hku@connect.hku.hk}
\affiliation{%
  \institution{Department of Electrical and Electronic Engineering, \\The University of Hong Kong}
  \city{HongKong SAR}
  \country{China}}

\author{Shuo Yang}
\email{shuoyang.ee@gmail.com}
\affiliation{%
  \institution{Department of Electrical and Electronic Engineering, \\The University of Hong Kong}
  \city{HongKong SAR}
  \country{China}}

\author{Hewei Wang}
\email{heweiw@andrew.cmu.edu}
\affiliation{%
    \institution{Robotics Institute, \\Carnegie Mellon University}  
    \city{Pittsburgh, PA}   
    \country{United States}}

\author{Yijie Li}
\email{yijieli@andrew.cmu.edu}
\affiliation{%
  \institution{Robotics Institute, \\Carnegie Mellon University}
  \city{Pittsburgh, PA}
  \country{United States}}

\author{Mengran Li}
\email{limr39@mail2.sysu.edu.cn}
\affiliation{%
  \institution{School of Intelligent Systems Engineering, \\Sun Yat-sen University}
  \city{Shenzhen}
  \country{China}}

\author{Puzhen Wu}
\email{puw4002@med.cornell.edu}
\affiliation{%
  \institution{Population Health Sciences, \\Weill Cornell Medicine}
  \city{New York, NY}
  \country{United States}}
    
\author{Edith C. H. Ngai}
\authornote{Corresponding authors}
\email{chngai@eee.hku.hk}
\affiliation{%
  \institution{Department of Electrical and Electronic Engineering, \\The University of Hong Kong}
  \city{HongKong SAR}
  \country{China}}

\renewcommand{\shortauthors}{Jinfeng Xu et al.}


 
\begin{abstract}
The data sparsity problem significantly hinders the performance of recommender systems, as traditional models rely on limited historical interactions to learn user preferences and item properties. While incorporating multimodal information can explicitly represent these preferences and properties, existing works often use it only as side information, failing to fully leverage its potential. In this paper, we propose MDVT, a model-agnostic approach that constructs multimodal-driven virtual triplets to provide valuable supervision signals, effectively mitigating the data sparsity problem in multimodal recommendation systems. To ensure high-quality virtual triplets, we introduce three tailored warm-up threshold strategies: static, dynamic, and hybrid. The static warm-up threshold strategy exhaustively searches for the optimal number of warm-up epochs but is time-consuming and computationally intensive. The dynamic warm-up threshold strategy adjusts the warm-up period based on loss trends, improving efficiency but potentially missing optimal performance. The hybrid strategy combines both, using the dynamic strategy to find the approximate optimal number of warm-up epochs and then refining it with the static strategy in a narrow hyper-parameter space. Once the warm-up threshold is satisfied, the virtual triplets are used for joint model optimization by our enhanced pair-wise loss function without causing significant gradient skew. Extensive experiments on multiple real-world datasets demonstrate that integrating MDVT into advanced multimodal recommendation models effectively alleviates the data sparsity problem and improves recommendation performance, particularly in sparse data scenarios.
\end{abstract}



\maketitle

\section{Introduction}
The rapid development of the internet has led to an information explosion, making recommender systems indispensable for navigating vast amounts of data. Traditional recommender systems rely on modeling user preferences through historical user-item interactions \cite{he2020lightgcn,xu2024fourierkan,xu2024aligngroup,chen2025squeeze}. However, the data sparsity problem significantly hinders the performance of these systems, as they depend solely on limited historical interactions to implicitly learn user preferences and item properties. Incorporating multimodal information \cite{chen2024rapverse,lin2025towards}—such as images and textual descriptions—allows for explicit representation of user preferences and item properties, potentially alleviating the data sparsity problem. Several recent works \cite{he2016vbpr,chen2017attentive,xu2025survey} have integrated multimodal content into recommendation models. For example, VBPR \cite{he2016vbpr} extends the matrix factorization framework to incorporate item visual features, while ACF \cite{chen2017attentive} introduces a hierarchically structured attention network to capture user preferences at the component level. Graph Convolutional Networks (GCNs) have also gained attention in this context \cite{guo2024lgmrec,wei2019mmgcn,wei2020graph,zhang2021mining,zhou2023tale}. Models like MMGCN \cite{wei2019mmgcn} and GRCN \cite{wei2020graph} employ GCNs to integrate multimodal information into the message-passing process, enhancing the inference of user and item representations. To further exploit the rich multimodal information, LATTICE \cite{zhang2021mining} and FREEDOM \cite{zhou2023tale} construct item-item graphs to aggregate semantically similar items. LGMRec \cite{guo2024lgmrec} utilizes hyper-graph structures to learn both global and local representations, capturing complex relationships in multimodal information. 

However, existing works typically use multimodal information only as side information to enhance the learning of user preferences, failing to fully leverage its potential. They primarily focus on improving item representations using multimodal content, while user representations are still learned solely from historical interactions. This limitation becomes more pronounced in data sparsity scenarios, where users have limited interaction records.

We propose that the similarity between user and item modality representations can serve as valuable supervision signals beyond explicit user-item interactions. To leverage this insight, we introduce \textbf{M}ultimodal-\textbf{D}riven \textbf{V}irtual \textbf{T}riplets (MDVT), a novel, model-agnostic approach that constructs virtual triplets based on multimodal information. These virtual triplets provide informative supervision signals, effectively mitigating the data sparsity problem in multimodal recommendation systems. A key challenge is that, unlike items, users do not have inherent multimodal information in recommendation scenarios. Users' modality representations must be learned from scratch by initializing embeddings randomly and refining them through model optimization. Consequently, the initial similarity between user and item modality representations may not provide high-quality supervision signals. To address this, we introduce three tailored warm-up threshold strategies:
\begin{itemize}[leftmargin=*]
    \item \textbf{Static Warm-up Threshold Strategy}: This strategy exhaustively searches for the optimal number of warm-up epochs, ensuring that user modality representations are sufficiently learned before constructing virtual triplets. While effective, it is time-consuming and computationally intensive due to the thorough hyper-parameter tuning required.
    \item \textbf{Dynamic Warm-up Threshold Strategy}: This strategy adjusts the warm-up period based on the trend of loss changes during training. It improves efficiency by reducing the need for extensive hyper-parameter tuning, automatically determining when user representations are adequately learned. However, it may not always find the optimal number of warm-up epochs compared to the static strategy.
    \item \textbf{Hybrid Warm-up Threshold Strategy}: Combining the strengths of both strategies, the hybrid strategy first employs the dynamic strategy to find the approximate optimal number of warm-up epochs and then applies the static strategy within a narrow hyper-parameter space. This allows for efficient training with a balance between computational cost and performance optimization.
\end{itemize}

Once the warm-up threshold is satisfied, the virtual triplets are used for joint model optimization through our enhanced pair-wise loss function, enhancing the learning process without causing significant gradient skew \cite{li2018visualizing,yu2020gradient}. Our MDVT approach is plug-and-play and can be easily integrated into any existing multimodal recommendation model, improving their performance, particularly in data sparsity scenarios. To validate the effectiveness of MDVT, we conducted extensive experiments on multiple real-world datasets adopting various advanced multimodal recommendation models. The results demonstrate that integrating MDVT into these models significantly alleviates the data sparsity problem and improves recommendation performance, especially for users with limited interaction records. Additionally, we would like to highlight the key distinction between our work and prior studies. Our virtual triplets are constructed based on the similarity between dynamically learned user and item representations, which are better aligned with the recommendation task with a sufficient warm-up phase. In contrast, prior works typically construct virtual samples based on the similarity of raw features between items.
\section{Preliminary}
In this section, we provide an overview of graph collaborative filtering (GCF), the common paradigm of advanced multimodal recommendations, which adopts graph neural network (GNN) into collaborative filtering (CF) with multimodal features. CF tasks usually contains a user set $\mathcal{U}=\{u_1,...,u_{|\mathcal{U}|}\}$ and an item set $\mathcal{I}=\{i_1,...,i_{|\mathcal{I}|}\}$. In multimodal scenarios, each item contains multiple features, we introduce modality-specific item embedding $i^m$ for each item $i$ belonging to the set of modalities $\mathcal{M}$. The user-item interaction matrix is denoted as $\mathcal{R} \in \{0,1\}^{|\mathcal{U}| \times |\mathcal{I}|}$. Specifically, each entry $\mathcal{R}_{u,i}$ indicates whether the user $u$ is connected to item $i$, with a value of 1 representing a connection and 0 otherwise. GCF naturally constructs the bipartite graph by user-item interaction matrix $\mathcal{R}$. This graph can be denoted by $\mathcal{G} = (\mathcal{U}, \mathcal{I}, \mathcal{E})$, where $\mathcal{U}, \mathcal{I}$ serve as the graph vertices, and $\mathcal{E}$ denotes the edge set. For each user-item pair $(u,i)$ that satisfies $\mathcal{R}_{u,i} = 1$, there exists bidirectional edges $(u,i) \in \mathcal{E}$ and $(i,u) \in \mathcal{E}$. We random initialize $\mathbf{E}_{u^m} \in \mathbb{R}^{d_m \times |\mathcal{U}|}$ to represent user embedding with modality $m$. $\mathbf{E}_{i^m} \in \mathbb{R}^{d_m \times|\mathcal{I}|}$ represents item initialized embedding with modality $m$, which extracted by pre-trained encoders. Here $d_m$ represents the hidden dimensionality. Based on the user-item graph $\mathcal{G}$, GNNs conduct neighbor aggregation to enhance user/item embeddings for extracting high-order user-item collaborative signals. Take the most widely-used GNN backbone LightGCN \cite{he2020lightgcn} as an example, the embeddings for user $u$ and item $i$ in the $l$-th layer are:
\begin{equation}
\label{eq:1}
\mathbf{e}_{u_{m}}^{(l)}=\frac{1}{d_u}\sum_{j | (u,j) \in \mathcal{E}} \frac{1}{d_j}\mathbf{e}_{j_{m}}^{(l-1)},  \quad \mathbf{e}_{i_{m}}^{(l)}=\frac{1}{d_i}\sum_{v |(i,v)\in\mathcal{E}} \frac{1}{d_v}\mathbf{e}_{v_{m}}^{(l-1)},
\end{equation}
where $d_{*}$ denotes degree of node. $\mathbf{e}_{*}^{(l)}$ represents node embedding in $l$-th layer. After $L$ layers of neighbor aggregation, the final representations of modality $m$ for user $u$ and item $i$ as:
\begin{equation}
\label{eq:2}
    \mathbf{\bar{e}}_{u_{m}} = \sum_{l=0}^{L} \mathbf{e}^{l}_{u_{m}}, \quad \mathbf{\bar{e}}_{i_{m}} = \sum_{l=0}^{L} \mathbf{e}^{l}_{i_{m}}.
\end{equation}

The predicted user-item relation score can be calculated by $\hat{y}_{u,i}=\sum_{m \in \mathcal{M}}(\bar{\mathbf{e}}_{u_{m}}^{\top} \bar{\mathbf{e}}_{i_{m}})$. With the prediction scores $\hat{y}_{u,i}$, the GNN models are optimized by
minimizing the BPR loss function \cite{rendle2009bpr}:
\begin{equation}
\label{eq:bpr}
    \mathcal{L}_{bpr} = \sum_{(u, i^{+}, i^{-}) \in \mathcal{D}} - \log(\sigma(\hat{y}_{u,i^{+}} - \hat{y}_{u,i^{-}})),
\end{equation}
where triplet training dataset $\mathcal{D}$ contains all positive user-item pairs $(u,i^{+}) \in \mathcal{E}$ and sampled negative user-item pairs $(u,i^{-}) \notin \mathcal{E}$. $\sigma(\cdot)$ denotes activation function. Though the above GCF paradigm achieves state-of-the-art performance in the recommendation field, its performance is limited by scarce interaction records. In light of this, this paper proposes MDVT, which leverages informative and valuable multimodal information to construct virtual training triplets to mitigate the data sparsity problem.
\section{Methodology}
\begin{figure*}
    \centering
    \includegraphics[width=0.95\linewidth]{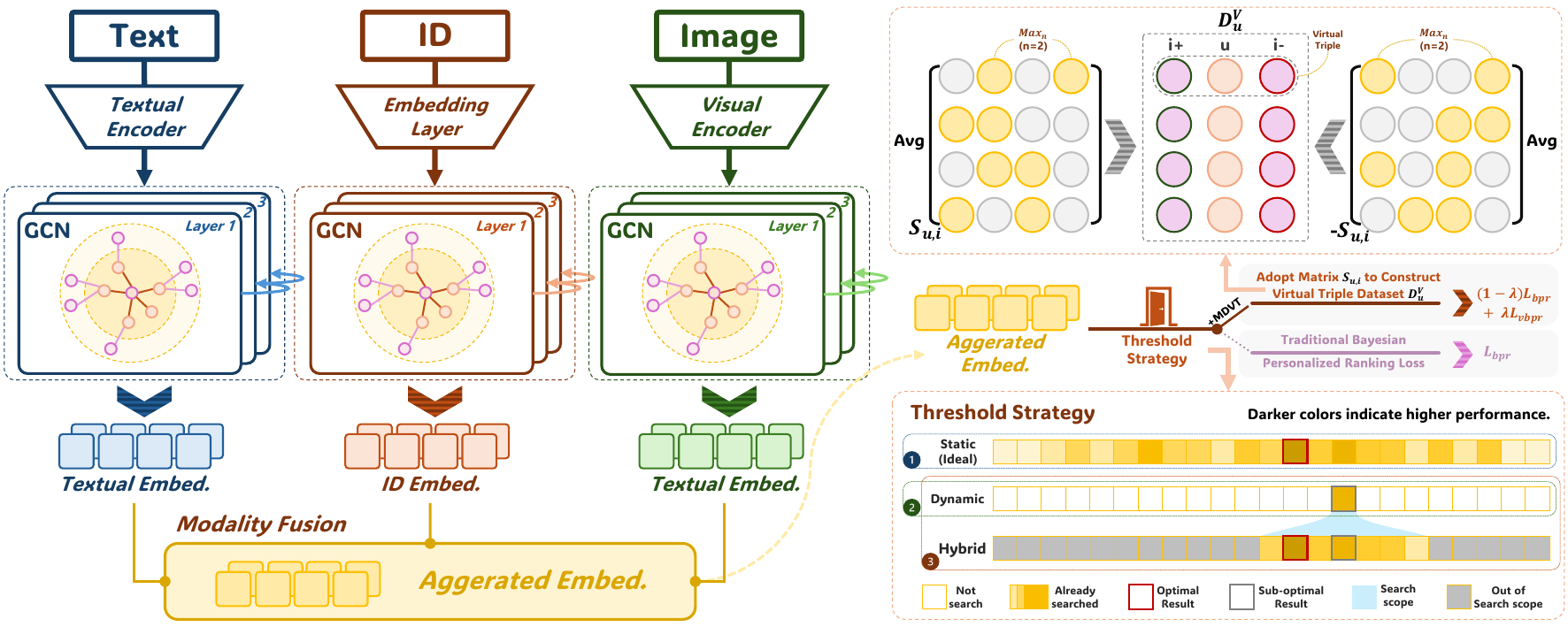}
    \vskip -0.15in
    \caption{The overall architecture of our proposed MDVT.}
    \label{fig:overview}
    \vskip -0.15in
\end{figure*}

In this section, we present our MDVT, a plug-and-play framework, which can improve all existing multimodal recommendation models' performance by constructing virtual training triplets to mitigate the data sparsity problem. The overall framework of our proposed MDVR is illustrated in Figure~\ref{fig:overview}\footnote{While Figure~\ref{fig:overview} only depicts ID embeddings, visual and textual modalities, it is important to note that our MDVT is model-agnostic and can be easily applied to all multimodal recommendation models, regardless of the number and types of modalities involved.}. Our proposed MDVT contains three main components:
\begin{itemize}[leftmargin=*]
    \item \textbf{Multimodal-Driven Virtual Triplets Constructor} (Section~\ref{sec:method1}): We construct virtual triplets by the top-$n$ positive and negative items for each user based on fused multimodal representation. 
    \item \textbf{Threshold Strategies} (Section~\ref{sec:method2}): To ensure the quality of virtual triples, we define three different threshold strategies: 1) \textbf{Static}: a heuristic static warm-up threshold strategy, 2) \textbf{Dynamic}: a loss-based dynamic warm-up threshold strategy, and 3) \textbf{Hybrid}: a hybrid warm-up threshold strategy.
    \item \textbf{Enhanced Pair-wise Loss Function} (Section~\ref{sec:method3}): Based on our constructed virtual triplets, we propose a simple yet effective enhanced pair-wise loss function, which can be directly plugged into all multimodal recommendation models.
\end{itemize}

\subsection{Multimodal-Driven Virtual Triplets}
\label{sec:method1}
Compared to traditional recommendation scenarios, items in multimodal recommendation settings contain rich modality features such as visual and textual information. While most previous multimodal recommendation studies \cite{jiang2024diffmm,zhou2023tale} have used multimodal information merely as side information to infer user preferences, recent studies \cite{zhou2024disentangled,ma2024xrec} in the explainable recommendation field demonstrate that leveraging multimodal data can explicitly reveal user preferences and item attributes. Inspired by these explainable recommendation approaches, we utilize modality information to provide additional supervision signals to alleviate the data sparsity problem that recommender systems often suffer from. To better plug into different multimodal recommendation models, we construct virtual triplets based on the final aggregated representations. We simplify the final representation aggregation for different multimodal recommendation models as:
\begin{equation}
\label{eq:4}
    \mathbf{\bar{e}}_u = \operatorname{F}(\mathbf{\bar{e}}_{u_m} | m \in \mathcal{M}) \quad \mathbf{\bar{e}}_i = \operatorname{F}(\mathbf{\bar{e}}_{i_m} | m \in \mathcal{M}), 
\end{equation}
where $\operatorname{F}(\cdot)$ represents modality representation fusion operation. Here, the fused representation contains valuable multimodal information \cite{zhou2023mmrec,zhou2023bootstrap,zhou2023tale}, which can be used to construct model-agnostic multimodal-driven virtual triplets. To calculate the top-$n$ positive and negative representation for each user node to construct virtual triplets, we maintain a user-item representation similarity matrix by cosine similarity, formally:
\begin{equation}
\label{eq:5}
    \mathcal{S}_{u,i} = \frac{\mathbf{\bar{e}}_u^{\top}\mathbf{\bar{e}_i}}{\|\mathbf{\bar{e}}_u\|\|\mathbf{\bar{e}_i}\|}.
\end{equation}

Based on the similarity matrix, for each user, we select the most similar $n$ and the least similar $n$ items to construct the virtual triplet: 
\begin{equation}
\label{eq:6}
    \mathcal{D}^{V}_{u,i^{+}} = \operatorname{Max_n}(\mathcal{S}_{u,i^{*}}|i^{*} \in \mathcal{I}), \quad
    \mathcal{D}^{V}_{u,i^{-}} = \operatorname{Max_n}(-\mathcal{S}_{u,i^{*}}|i^{*} \in \mathcal{I}),
\end{equation}
where $\operatorname{Max_n}(\cdot)$ denotes top-$n$ similarity filter operation. For user $u$, $\mathcal{D}^{V}_{u,i^{+}}$ and $\mathcal{D}^{V}_{u,i^{-}}$ contain $n$ positive items and negative items, respectively. Therefore, we construct a new virtual triplet training dataset $\mathcal{D}^{V}$, which is updated with model optimization. For each triplet with user $u$ in this virtual triplet training dataset $\mathcal{D}^{V}$ can be expressed as $\mathcal{D}^V_u = (u, \mathcal{D}^{V}_{u,i^{+}}, \mathcal{D}^{V}_{u,i^{-}})$. 

\subsection{Threshold Strategies}
\label{sec:method2}
High-quality representations are essential for constructing effective virtual triplets. It is worth noting that, unlike items—which naturally possess multimodal information such as images and textual descriptions—users in recommendation scenarios do not inherently have multimodal information. Therefore, user representations are initially randomized and progressively refined during training, the model requires sufficient warm-up epochs to ensure the representations are adequately optimized for constructing high-quality triplets. To this end, we propose three threshold strategies to determine the optimal number of warm-up epochs. These strategies ensure that virtual triplets are incorporated only when the representation quality is sufficient to provide high-quality supervision signals. Specifically, we propose three warm-up threshold strategies: Static Warm-up Threshold (\textbf{Static}), Dynamic Warm-up Threshold (\textbf{Dynamic}), and Hybrid Warm-up Threshold (\textbf{Hybrid}).

\subsubsection{Static Warm-up Threshold Strategy.}
The warm-up epochs required for learning high-quality representations varies across different models and hyper-parameter settings (e.g., learning rate, batch size). Therefore, a simple and effective strategy is to set a pre-defined threshold $\mathcal{T_S}$, and our virtual triplets will jointly optimize the multimodal recommendation model after $\mathcal{T_S}$ epochs training. For our static warm-up threshold strategy, we search the optimal number of warm-up epochs within a manually defined threshold set $\mathcal{S_T}$. When computational resources are abundant, we can search for the optimal parameters by exhaustively traversing the hyper-parameter space. Conversely, when computational resources are limited, we rely on researchers' extensive domain knowledge and experience in model training to define threshold set $\mathcal{S_T}$. 



\subsubsection{Dynamic Warm-up Threshold Strategy.}
In scenarios with limited computational resources, exhaustively traversing all hyper-parameters to find the optimal warm-up epoch number is impractical, and researchers may not be familiar with all existing models to manually set thresholds appropriately. Therefore, a dynamic threshold strategy with a lower hyper-parameter tuning cost is necessary. Inspired by numerous studies \cite{sutskever2013importance,li2018visualizing,bottou2018optimization,wilson2017marginal} on the relationship between training loss and model convergence, we propose a dynamic warm-up threshold strategy based on the trend of loss changes. Specifically, we assess whether the model is approaching convergence by comparing the ratio of the loss decrease between the current and previous epochs. When the rate of loss change is low enough, indicating that the model has sufficiently converged and this epoch is an approximate optimal threshold $\mathcal{T_{O}}$. Then, we adopt virtual triplets to optimize the multimodal recommendation model jointly. We define the loss at epoch $t$ as $\mathcal{L}^{t}$. Virtual triplets are incorporated into the model optimization process when the rate of loss change falls below a pre-defined hyper-parameter $g$.



\subsubsection{Hybrid Warm-up Threshold Strategy.}
An effective and satisfactory threshold selection strategy is the hybrid warm-up threshold strategy, which combines the dynamic warm-up threshold strategy and static warm-up threshold strategy. Specifically, it adopts the dynamic warm-up threshold strategy to find an approximate optimal threshold $\mathcal{T}^{cur}$, then adopts the static warm-up threshold strategy within a small scope $[\mathcal{T}^{cur}-s, \mathcal{T}^{cur}+s]$ to find the optimal threshold $\mathcal{T_{O}}$, where $s$ is the search scope hyper-parameter. This hybrid warm-up threshold strategy allows for a high probability of finding the optimal number of warm-up epochs without searching the entire hyper-parameter space. 


\textbf{Analysis.} The static warm-up threshold strategy requires comprehensive hyper-parameter tuning because it involves manually selecting the optimal number of warm-up epochs through an exhaustive search of the hyper-parameter space. In scenarios where a full traversal of hyper-parameters is feasible, this strategy can effectively find the optimal number of warm-up epochs, ensuring the model performs at its best. However, this process is time-consuming and computationally intensive due to the high demand for hyper-parameter tuning. In contrast, the dynamic warm-up threshold strategy reduces the need for extensive hyper-parameter tuning by automatically selecting the number of warm-up rounds based on the trend of loss change. This strategy adjusts dynamically to each model's loss change trend, allowing for a more efficient training process with lower hyper-parameter tuning demands. The dynamic warm-up threshold strategy is particularly beneficial when computational resources are limited. However, despite its advantages, the dynamic warm-up threshold strategy may not always find the optimal number of warm-up epochs compared to the static warm-up threshold strategy with full hyper-parameter traversal. Since it does not exhaustively explore the hyper-parameter space, there's a possibility that it might miss the optimal number of warm-up epochs for a given model. Therefore, while the dynamic warm-up threshold strategy improves efficiency and requires less manual tuning, it might sacrifice some performance optimization achievable through the static method. Moreover, the hybrid warm-up threshold strategy combines the advantages of both the static and dynamic warm-up threshold strategies. Specifically, it first adopts the dynamic warm-up threshold strategy to find the approximate optimal number of warm-up epochs. Then, it applies the static warm-up threshold strategy within a small scope, offering the potential for optimal performance. We present the procedure in Appendix~\ref{appendix:algorithm}.

\subsection{Enhanced Pair-wise Loss Function}
\label{sec:method3}
Once the multimodal recommendation model has learned high-quality representations (when the threshold strategy is satisfied), we adopt the widely used Bayesian Personalized Ranking (BPR) loss on our virtual triplet training dataset to optimize the model:
\begin{equation}
\label{eq:vbpr}
    \mathcal{L}_{vbpr} = \sum_{(u, \mathcal{D}^{V}_{u,i^{+}}, \mathcal{D}^{V}_{u,i^{-}}) \in \mathcal{D}^V} - \log(\sigma(\mathbf{\bar{e}}_u^{\top}\mathbf{\hat{e}}_{u,i^{+}} - \mathbf{\bar{e}}_u^{\top}\mathbf{\hat{e}}_{u,i^{-}})),
\end{equation}
\begin{equation}
\label{eq:agr}
    \mathbf{\hat{e}}_{u,i^{+}} = \frac{1}{n}\sum_{i^{+} \in \mathcal{D}^{V}_{u,i^{+}}}\mathbf{\bar{e}}_{i^{+}} \quad  \mathbf{\hat{e}}_{u,i^{-}} = \frac{1}{n}\sum_{i^{-} \in \mathcal{D}^{V}_{u,i^{-}}}\mathbf{\bar{e}}_{i^{-}},
\end{equation}
where we calculate the representation mean of the similar group and the representation mean of the dissimilar group to get informative representations of the virtual positive item and negative item, respectively. We jointly optimize the model with two loss functions: the BPR loss defined in Eq.\ref{eq:bpr} applied to the training dataset $\mathcal{D}$, and the BPR loss applied to the virtual training dataset $\mathcal{D}^V$ as defined in Eq.\ref{eq:vbpr}. The final learning loss can be expressed as:
\begin{equation}
\label{eq:loss}
    \mathcal{L} = (1-\lambda)\mathcal{L}_{bpr} + \lambda \mathcal{L}_{vbpr},
\end{equation}
where $\lambda$ regulates the influence of our virtual training loss $\mathcal{L}_{vbpr}$. Note that the users involved in $\mathcal{L}_{bpr}$ and $\mathcal{L}_{vbpr}$ completely overlap, and each training triplet in both cases consists of one positive item and one negative item. Adding $\mathcal{L}_{vbpr}$ alters the loss magnitude, so $\lambda$ ensures balanced scaling to prevent gradient skew \cite{li2018visualizing,yu2020gradient}, enabling smooth joint optimization.

\section{Evaluation}
\label{sec:eval}
We conduct extensive experiments on MDVT, aiming to answer the following research questions (RQs): \textbf{RQ1:} Can MDVT enhance the performance of multimodal recommender systems? \textbf{RQ2:} How do the various components in MDVT affect performance enhancement? \textbf{RQ3:} What are the effectiveness and costs of different threshold strategies in MDVT? \textbf{RQ4:} Can MDVT have a positive impact on the convergence speed? \textbf{RQ5:} Can MDVT be compatible with robust training and data augmentation strategies? \textbf{RQ6:} How do different warm-up threshold strategies work in practical training? \textbf{RQ7:} What is the impact of key hyper-parameters in MDVT?

\subsection{Experimental Settings}
\subsubsection{Datasets}
The experiments are conducted on three real-world datasets: Baby, Sports, and Clothing from Amazon \cite{mcauley2015image}. All the datasets comprise textual and visual features in the form of item descriptions and images. To further evaluate the performance of MDVT in scenarios involving multiple modalities, we also conduct experiments on the TikTok dataset \cite{jiang2024diffmm}. Our data preprocessing methodology follows the approach outlined in MMRec \cite{zhou2023mmrecsm}. Table~\ref{tab:dataset_statistics} shows the statistics of these datasets. We adopt two widely used metrics to evaluate the performance fairly: Recall@K (R@K) and NDCG@K (N@K). We report the average metrics of all users in the test dataset under both K = 5 and K = 10. We follow the popular evaluation setting \cite{zhou2023tale,guo2024lgmrec} with a random data splitting 8:1:1 for training, validation, and testing.

\subsubsection{Baselines}
We extensively examine the performance of our MDVT across a variety of multimodal recommendation models, including MMGCN \cite{wei2019mmgcn}, SLMRec \cite{tao2022self}, FREEDOM \cite{zhou2023tale}, DRAGON \cite{zhou2023bootstrap}, LGMRec \cite{guo2024lgmrec}, and MMSSL \cite{wei2023multi}. Moreover, we test the compatibility of our MDVT with the adversarial training strategy (AMR \cite{tang2019adversarial}) and LLM-based data augmentation strategy (GPT-4o \cite{yang2023dawn}).

\begin{table}[!t]
 \vskip -0.05in
    \centering
\caption{Statistics of the three evaluation datasets.}
\small
\setlength{\tabcolsep}{1.0mm}
 \vskip -0.15in
\label{tab:dataset_statistics}
    \begin{tabular}{cccccc}
    \toprule
         \textbf{Datasets}&  \textbf{\#Users}&  \textbf{\#Items}& \textbf{\#Interactions} & \textbf{Sparsity} & \textbf{Modality}\\
         \midrule
         Baby & 19,445 & 7,050 & 160,792 & 99.88\% & V,T\\
         Sports & 35,598 & 18,357 & 296,337 & 99.95\% & V,T\\
         Clothing & 39,387 & 23,033 & 278,677 & 99.97\% & V,T\\
         TikTok & 9,319 & 6,710 & 59,541 & 99.90\% & V,T,A\\
         \bottomrule
    \end{tabular}
     \vskip -0.15in
\end{table}

\begin{table*}[!t]
\caption{Performance comparison of baselines with or without MDVT on all datasets in terms of Recall@K (R@K) and NDCG@K (N@K). $^*$ indicates the improvement is statistically significant, where the p-value is less than 0.01. (S), (D), and (H) denote Static, Dynamic, and Hybrid, respectively.}
  \vskip -0.15in
\centering
\tabcolsep=0.04in
\label{tab:comparison results}
\resizebox{\linewidth}{!}{
    \begin{tabular}{c|cccc|cccc|cccc|cccc}
     \toprule
         Datasets&  \multicolumn{4}{c|}{Baby}&  \multicolumn{4}{c|}{Sports}&  \multicolumn{4}{c|}{Clothing}&  \multicolumn{4}{c}{TikTok}\\\midrule
         Metrics& R@5& R@10& N@5& N@10 & R@5& R@10& N@5& N@10& R@5& R@10& N@5& N@10& R@5& R@10& N@5& N@10\\\midrule
         MMGCN & 0.0240& 0.0378& 0.0160& 0.0200& 0.0216& 0.0370& 0.0143& 0.0193& 0.0130& 0.0218& 0.0088& 0.0110& 0.0331& 0.0463& 0.0172& 0.0231\\
         +MDVT (S) & \underline{0.0255$^*$}& \underline{0.0418$^*$}& \underline{0.0169$^*$}& \underline{0.0221$^*$}& \underline{0.0234$^*$}& \underline{0.0404$^*$}& \underline{0.0157$^*$}& \underline{0.0211$^*$}& \textbf{0.0148$^*$}& \textbf{0.0244$^*$}& \textbf{0.0099$^*$}& \textbf{0.0126$^*$}& \underline{0.0381$^*$}& \underline{0.0539$^*$}& \underline{0.0201$^*$}& \underline{0.0268}$^*$\\
         \gray Improv. &\gray 6.25\%&\gray 10.58\%&\gray 5.62\%&\gray 10.50\%&\gray 8.33\%&\gray 9.19\%&\gray 9.79\%&\gray 9.33\%&\gray 13.85\%&\gray 11.93\%&\gray 12.50\%&\gray 14.55\%&\gray 15.11\%&\gray 16.41\%&\gray 16.86\%&\gray 16.02\%\\
         +MDVT (D) & 0.0251$^*$& 0.0413$^*$& 0.0167$^*$& 0.0218$^*$& 0.0231$^*$& 0.0400$^*$& 0.0155$^*$& 0.0208$^*$& \textbf{0.0148$^*$}& \textbf{0.0244$^*$}& \textbf{0.0099$^*$}& \textbf{0.0126$^*$}& 0.0373$^*$& 0.0520$^*$& 0.0197$^*$& 0.0263$^*$\\
         \gray Improv. &\gray 4.58\%&\gray 9.26\%&\gray 4.37\%&\gray 9.00\%&\gray 6.94\%&\gray 8.11\%&\gray 8.39\%&\gray 7.77\%&\gray 13.85\%&\gray 11.93\%&\gray 12.50\%&\gray 14.55\%&\gray 12.69\%&\gray 12.31\%&\gray 14.53\%&\gray 13.85\%\\
         +MDVT (H) & \textbf{0.0257$^*$}& \textbf{0.0420$^*$}& \textbf{0.0170$^*$}& \textbf{0.0224$^*$}& \textbf{0.0236$^*$}& \textbf{0.0406$^*$}& \textbf{0.0158$^*$}& \textbf{0.0213$^*$}& \textbf{0.0148$^*$}& \textbf{0.0244$^*$}& \textbf{0.0099$^*$}& \textbf{0.0126$^*$}& \textbf{0.0383$^*$}& \textbf{0.0543$^*$}& \textbf{0.0203$^*$}& \textbf{0.0272$^*$}\\
         \gray Improv. &\gray 7.08\%&\gray 11.11\%&\gray 6.25\%&\gray 12.00\%&\gray 9.26\%&\gray 9.73\%&\gray 10.49\%&\gray 10.36\%&\gray 13.85\%&\gray 11.93\%&\gray 12.50\%&\gray 14.55\%&\gray 15.71\%&\gray 17.28\%&\gray 18.02\%&\gray 17.75\%\\\midrule
         SLMRec & 0.0343& 0.0529& 0.0226& 0.0290& 0.0429& 0.0663& 0.0288& 0.0365& 0.0292& 0.0452& 0.0196& 0.0247& 0.0349& 0.0503& 0.0188& 0.0251\\
         +MDVT (S) & \underline{0.0357$^*$}& \underline{0.0560$^*$}& \underline{0.0239$^*$}& \underline{0.0319$^*$}& \underline{0.0458$^*$}& \underline{0.0705$^*$}& \underline{0.0307$^*$}& \underline{0.0388$^*$}& \underline{0.0317$^*$}& \underline{0.0493$^*$}& \underline{0.0214$^*$}& \underline{0.0267$^*$}& \underline{0.0384$^*$}& \underline{0.0550$^*$}& \underline{0.0208$^*$}& \underline{0.0276$^*$}\\
         \gray Improv. &\gray 4.08\%&\gray 5.86\%&\gray 5.75\%&\gray 10.00\%&\gray 6.76\%&\gray 6.33\%&\gray 6.60\%&\gray 6.30\%&\gray 8.56\% &\gray 9.07\%&\gray 9.18\%&\gray 8.10\%&\gray 10.03\%&\gray 9.34\%&\gray 10.64\%&\gray 9.96\%\\
         +MDVT (D) & 0.0353$^*$& 0.0553$^*$& 0.0234$^*$& 0.0313$^*$& 0.0456$^*$& 0.0701$^*$& 0.0305$^*$& 0.0385$^*$& 0.0314$^*$& 0.0488$^*$& 0.0212$^*$& 0.0263$^*$& 0.0379$^*$& 0.0542$^*$& 0.0204$^*$& 0.0271$^*$\\
         \gray Improv. &\gray 2.92\%&\gray 4.54\%&\gray 3.54\%&\gray 7.93\%&\gray 6.29\%&\gray 5.73\%&\gray 5.90\%&\gray 5.48\%&\gray 7.53\% &\gray 7.96\%&\gray 8.16\%&\gray 6.48\%&\gray 8.60\%&\gray 7.75\%&\gray 8.51\%&\gray 7.97\%\\
         +MDVT (H) & \textbf{0.0359$^*$}& \textbf{0.0563$^*$}& \textbf{0.0241$^*$}& \textbf{0.0323$^*$}& \textbf{0.0460$^*$}& \textbf{0.0709$^*$}& \textbf{0.0309$^*$}& \textbf{0.0391$^*$}& \textbf{0.0319$^*$}& \textbf{0.0497$^*$}& \textbf{0.0216$^*$}& \textbf{0.0270$^*$}& \textbf{0.0393$^*$}& \textbf{0.0566$^*$}& \textbf{0.0209$^*$}& \textbf{0.0284$^*$}\\
         \gray Improv. &\gray 4.66\%&\gray 6.43\%&\gray 6.64\%&\gray 11.38\%&\gray 7.23\%&\gray 6.94\%&\gray 7.29\%&\gray 7.12\%&\gray 9.25\% &\gray 9.96\%&\gray 10.20\%&\gray 9.31\%&\gray 12.61\%&\gray 12.52\%&\gray 11.17\%&\gray 13.15\%\\\midrule
         FREEDOM & 0.0374& 0.0627& 0.0243& 0.0330& 0.0446& 0.0717& 0.0291& 0.0385& 0.0388& 0.0629& 0.0257& 0.0341& 0.0399& 0.0589& 0.0214& 0.0295\\
         +MDVT (S) & \underline{0.0391$^*$}& \underline{0.0652$^*$}& \underline{0.0257$^*$}& \underline{0.0350$^*$}& \underline{0.0472$^*$}& \underline{0.0752$^*$}& \underline{0.0312$^*$}& \underline{0.0406$^*$}& \underline{0.0410$^*$}& \underline{0.0662$^*$}& \underline{0.0275$^*$}& \underline{0.0361$^*$}& \underline{0.0427$^*$}& \underline{0.0629$^*$}& \underline{0.0226$^*$}& \underline{0.0312$^*$}\\
         \gray Improv. &\gray 4.55\%&\gray 3.99\%&\gray 5.76\%&\gray 6.06\%&\gray 5.83\%&\gray 4.88\%&\gray 7.22\%&\gray 5.45\%&\gray 5.67\%&\gray 5.25\%&\gray 7.00\%&\gray 5.87\%&\gray 7.02\%&\gray 6.79\%&\gray 5.61\%&\gray 5.76\%\\
         +MDVT (D) & 0.0387$^*$& 0.0648$^*$& 0.0253$^*$& 0.0347$^*$& 0.0469$^*$& 0.0750$^*$& 0.0310$^*$& 0.0403$^*$& 0.0405$^*$& 0.0655$^*$& 0.0270$^*$& 0.0354$^*$& 0.0420$^*$& 0.0622$^*$& 0.0223$^*$& 0.0309$^*$\\
         \gray Improv. &\gray 3.48\%&\gray 3.35\%&\gray 4.12\%&\gray 5.15\%&\gray 5.16\%&\gray 4.60\%&\gray 6.53\%&\gray 4.68\%&\gray 4.38\%&\gray 4.13\%&\gray 5.06\%&\gray 3.81\%&\gray 5.26\%&\gray 5.60\%&\gray 4.21\%&\gray 4.75\%\\
         +MDVT (H) & \textbf{0.0398$^*$}& \textbf{0.0662$^*$}& \textbf{0.0262$^*$}& \textbf{0.0357$^*$}& \textbf{0.0476$^*$}& \textbf{0.0757$^*$}& \textbf{0.0315$^*$}& \textbf{0.0410$^*$}& \textbf{0.0412$^*$}& \textbf{0.0665$^*$}& \textbf{0.0277$^*$}& \textbf{0.0364$^*$}& \textbf{0.0431$^*$}& \textbf{0.0629$^*$}& \textbf{0.0229$^*$}& \textbf{0.0318$^*$}\\
         \gray Improv. &\gray 6.42\%&\gray 5.58\%&\gray 7.82\%&\gray 8.18\%&\gray 6.73\%&\gray 5.58\%&\gray 8.25\%&\gray 6.49\%&\gray 6.19\%&\gray 5.72\%&\gray 7.78\%&\gray 6.74\%&\gray 8.02\%&\gray 6.79\%&\gray 7.01\%&\gray 7.80\%\\
         \midrule
        
         DRAGON & 0.0380& 0.0662& 0.0249& 0.0345& 0.0449& 0.0752& 0.0296& 0.0413& 0.0401& 0.0671& 0.0270& 0.0365& 0.0451& 0.0682& 0.0244& 0.0341\\
         +MDVT (S) & \underline{0.0396$^*$}& \underline{0.0689$^*$}& \underline{0.0262$^*$}& \underline{0.0360$^*$}& \underline{0.0474$^*$}& \underline{0.0780$^*$}& \underline{0.0311$^*$}& \underline{0.0434$^*$}& \underline{0.0430$^*$}& \underline{0.0710$^*$}& \underline{0.0287$^*$}& \underline{0.0385$^*$}& \underline{0.0475$^*$}& \underline{0.0718$^*$}& \underline{0.0259$^*$}& \underline{0.0362$^*$}\\
         \gray Improv. &\gray 4.21\%&\gray 4.08\%&\gray 5.22\%&\gray 4.35\%&\gray 5.57\%&\gray 3.72\%&\gray 5.07\%&\gray 5.08\%&\gray 7.23\%&\gray 5.81\%&\gray 6.30\%&\gray 5.48\%&\gray 5.32\%&\gray 5.28\%&\gray 6.15\%&\gray 6.16\%\\
         +MDVT (D) & 0.0391$^*$& 0.0685$^*$& 0.0259$^*$& 0.0357$^*$& 0.0470$^*$& 0.0776$^*$& 0.0308$^*$& 0.0430$^*$& 0.0428$^*$& 0.0704$^*$& 0.0285$^*$& 0.0382$^*$& 0.0471$^*$& 0.0712$^*$& 0.0257$^*$& 0.0359$^*$\\
         \gray Improv. &\gray 2.89\%&\gray 3.47\%&\gray 4.02\%&\gray 3.48\%&\gray 4.68\%&\gray 3.20\%&\gray 4.05\%&\gray 4.12\%&\gray 6.73\%&\gray 4.92\%&\gray 5.56\%&\gray 4.66\%&\gray 4.43\%&\gray 4.40\%&\gray 5.33\%&\gray 5.28\%\\
         +MDVT (H) & \textbf{0.0398$^*$}& \textbf{0.0692$^*$}& \textbf{0.0264$^*$}& \textbf{0.0364$^*$}& \textbf{0.0479$^*$}& \textbf{0.0788$^*$}& \textbf{0.0314$^*$}& \textbf{0.0440$^*$}& \textbf{0.0432$^*$}& \textbf{0.0713$^*$}& \textbf{0.0288$^*$}& \textbf{0.0387$^*$}& \textbf{0.0480$^*$}& \textbf{0.0724$^*$}& \textbf{0.0262$^*$}& \textbf{0.0366$^*$}\\
         \gray Improv. &\gray 4.74\%&\gray 4.53\%&\gray 6.02\%&\gray 5.51\%&\gray 6.68\%&\gray 4.79\%&\gray 6.08\%&\gray 6.54\%&\gray 7.73\%&\gray 6.26\%&\gray 6.67\%&\gray 6.03\%&\gray 6.43\%&\gray 6.16\%&\gray 7.38\%&\gray 7.33\%\\\midrule
         LGMRec & 0.0374& 0.0626& 0.0249& 0.0333& 0.0446& 0.0719& 0.0288& 0.0387& 0.0371& 0.0555& 0.0246& 0.0302& 0.0406& 0.0610& 0.0217& 0.0304\\
         +MDVT (S) & \underline{0.0416$^*$}& \underline{0.0656$^*$}& \underline{0.0280$^*$}& \underline{0.0359$^*$}& \underline{0.0474$^*$}& \underline{0.0769$^*$}& \underline{0.0311$^*$}& \underline{0.0417$^*$}& \underline{0.0409$^*$}& \underline{0.0619$^*$}& \underline{0.0274$^*$}& \underline{0.0335$^*$}& \underline{0.0431$^*$}& \underline{0.0641$^*$}& \underline{0.0231$^*$}& \underline{0.0320$^*$}\\
         \gray Improv. &\gray 11.23\%&\gray 4.79\%&\gray 12.45\%&\gray 7.81\%&\gray 6.28\%&\gray 6.95\%&\gray 7.99\%&\gray 7.75\%&\gray 10.24\%&\gray 11.53\%&\gray 11.38\%&\gray 10.93\%&\gray 6.16\%&\gray 5.08\%&\gray 6.45\%&\gray 5.26\%\\
         +MDVT (D) & 0.0413$^*$& 0.0651$^*$& 0.0279$^*$& 0.0356$^*$& 0.0471$^*$& 0.0762$^*$& 0.0308$^*$& 0.0412$^*$& 0.0403$^*$& 0.0612$^*$& 0.0270$^*$& 0.0328$^*$& 0.0425$^*$& 0.0635$^*$& 0.0228$^*$& 0.0317$^*$\\
         \gray Improv. &\gray 10.43\%&\gray 3.99\%&\gray 12.05\%&\gray 6.91\%&\gray 5.61\%&\gray 5.98\%&\gray 6.94\%&\gray 6.46\%&\gray 8.63\%&\gray 10.27\%&\gray 9.76\%&\gray 8.61\%&\gray 4.68\%&\gray 4.10\%&\gray 5.07\%&\gray 4.28\%\\
         +MDVT (H) & \textbf{0.0417$^*$}& \textbf{0.0660$^*$}& \textbf{0.0281$^*$}& \textbf{0.0360$^*$}& \textbf{0.0475$^*$}& \textbf{0.0771$^*$}& \textbf{0.0312$^*$}& \textbf{0.0419$^*$}& \textbf{0.0411$^*$}& \textbf{0.0622$^*$}& \textbf{0.0276$^*$}& \textbf{0.0337$^*$}& \textbf{0.0435$^*$}& \textbf{0.0647$^*$}& \textbf{0.0233$^*$}& \textbf{0.0324$^*$}\\
         \gray Improv. &\gray 11.50\%&\gray 5.43\%&\gray 12.85\%&\gray 8.11\%&\gray 6.50\%&\gray 7.23\%&\gray 8.33\%&\gray 8.27\%&\gray 10.78\%&\gray 12.07\%&\gray 12.20\%&\gray 11.92\%&\gray 7.14\%&\gray 6.07\%&\gray 7.37\%&\gray 6.58\%\\\midrule
         MMSSL & 0.0369& 0.0613& 0.0241& 0.0326& 0.0451& 0.0693& 0.0294& 0.0369& 0.0382& 0.0619& 0.0253& 0.0335& 0.0395& 0.0575& 0.0210& 0.0287\\
         +MDVT (S) & \underline{0.0396$^*$}& \underline{0.0655$^*$}& \underline{0.0257$^*$}& \underline{0.0348$^*$}& \underline{0.0478$^*$}& \underline{0.0732$^*$}& \underline{0.0312$^*$}& \underline{0.0389$^*$}& \underline{0.0402$^*$}& \underline{0.0648$^*$}& \underline{0.0267$^*$}& \underline{0.0351$^*$}& \underline{0.0423$^*$}& \underline{0.0613$^*$}& \underline{0.0224$^*$}& \underline{0.0305$^*$}\\
         \gray Improv. &\gray 7.32\%&\gray 6.85\%&\gray 6.64\%&\gray 6.75\%&\gray 5.99\%&\gray 5.63\%&\gray 6.12\%&\gray 5.42\%&\gray 5.24\%&\gray 4.68\%&\gray 5.53\%&\gray 4.78\%&\gray 7.09\%&\gray 6.61\%&\gray 6.67\%&\gray 6.27\%\\
         +MDVT (D) & 0.0388$^*$& 0.0649$^*$& 0.0252$^*$& 0.0343$^*$& 0.0473$^*$& 0.0727$^*$& 0.0308$^*$& 0.0385$^*$& 0.0400$^*$& 0.0643$^*$& 0.0263$^*$& 0.0348$^*$& 0.0419$^*$& 0.0609$^*$& 0.0221$^*$& 0.0302$^*$\\
         \gray Improv. &\gray 5.15\%&\gray 5.87\%&\gray 4.56\%&\gray 5.21\%&\gray 4.88\%&\gray 4.91\%&\gray 4.76\%&\gray 4.34\%&\gray 4.71\%&\gray 3.88\%&\gray 3.95\%&\gray 3.88\%&\gray 6.08\%&\gray 5.91\%&\gray 5.24\%&\gray 5.23\%\\
         +MDVT (H) & \textbf{0.0404$^*$}& \textbf{0.0667$^*$}& \textbf{0.0262$^*$}& \textbf{0.0354$^*$}& \textbf{0.0486$^*$}& \textbf{0.0745$^*$}& \textbf{0.0317$^*$}& \textbf{0.0396$^*$}& \textbf{0.0407$^*$}& \textbf{0.0657$^*$}& \textbf{0.0270$^*$}& \textbf{0.0356$^*$}& \textbf{0.0429$^*$}& \textbf{0.0621$^*$}& \textbf{0.0228$^*$}& \textbf{0.0310$^*$}\\
         \gray Improv. &\gray 9.49\%&\gray 8.81\%&\gray 8.71\%&\gray 8.59\%&\gray 7.76\%&\gray 7.50\%&\gray 7.82\%&\gray 7.32\%&\gray 6.54\%&\gray 6.14\%&\gray 6.72\%&\gray 6.27\%&\gray 8.61\%&\gray 8.00\%&\gray 8.57\%&\gray 8.01\%\\
        \bottomrule
    \end{tabular}
    }
    \vskip -0.15in
\end{table*}

\subsubsection{Implementation Details}
\label{sec:impl}
We retain the standard settings for all baselines and fix batch size $B$ to 2048. For MDVT, we apply a grid search on hyper-parameters $\lambda$ in $\{0.1, 0.2, 0.3, 0.4, 0.5\}$, and the value $n$ in Eq.~\ref{eq:6} in $\{1, 2, 4, 8\}$. For the static warm-up threshold strategy, we define threshold set as $\mathcal{S_T} = \{0, 5, 10, 20, 40, 80\}$. We perform a grid search in $\{0.1, 0.2, 0.3, 0.4\}$ for hyper-parameter $g$ in dynamic and hybrid warm-up threshold strategies, a grid search in $\{1, 2, 3, 4, 5\}$ for hyper-parameter $s$ in hybrid warm-up threshold strategy. The common optimizer is Adam \cite{kingma2014adam} and all training and evaluation of all models are conducted on an RTX4090 GPU. For the GPT-4o data augmentation strategy, we utilize GPT-4o \cite{yang2023dawn} to augment the raw description via the items' image for all datasets to improve the correlation between textual and visual modalities. We designed the prompt as: \textbf{`[$V$] Here is the description of an item and the corresponding picture, please combine the picture to improve the description quality in one paragraph. The description is as follows: [$T$].'}, where [$V$] and [$T$] are the raw image and description for each item, respectively.

\subsection{Overall Performance (RQ1)}
\label{subsec:RQ1}
We evaluate the effectiveness of our MDVT with various warm-up threshold strategies on various models for multimodal recommendation scenarios. From Table~\ref{tab:comparison results}, we find the following observations:

\noindent \textbf{\underline{Observation1}: MDVT with all warm-up threshold strategies can enhance the performance of various multimodal recommendation models.} As shown in Table~\ref{tab:comparison results}, we conduct extensive experiments with MDVT on five multimodal recommendation models across three distinct public datasets. The experimental results demonstrate that all warm-up threshold strategies significantly improve over all baselines across all evaluation metrics. The static warm-up threshold strategy consistently achieves superior results compared to the dynamic warm-up threshold strategy. Moreover, the hybrid warm-up threshold strategy consistently outperforms both the static and dynamic warm-up threshold strategies. In summary, the experimental results validate that leveraging multimodal information to construct virtual triplets can effectively improve recommender performance by mitigating the data sparsity problem.

\noindent \textbf{\underline{Observation2}: Hybrid warm-up threshold strategy can find ideal warm-up epochs within an affordable hyper-parameter tuning cost.} For all multimodal recommendation models across all datasets, the dynamic warm-up threshold strategy achieves performance improvements comparable to the static warm-up threshold strategy. This indicates that the dynamic warm-up threshold strategy can identify approximately optimal warm-up epochs without requiring extensive hyper-parameter tuning or substantial expert knowledge. Building on this, the hybrid warm-up threshold strategy utilizes the approximately optimal number of warm-up epochs found by the dynamic strategy to adopt the static warm-up threshold strategy within a small scope. Consequently, it finds the ideal number of warm-up epochs closer to the optimal number and achieves superior performance improvements within an affordable hyper-parameter tuning cost.

\subsection{Ablation Study (RQ2)}
To discern the impact of our MDVT's core components, we conducted an ablation study with various configurations:
\begin{itemize}[leftmargin=*]
    \item $w/o$-Aggr: This configuration removes the representation averaging operation specified in Eq.~\ref{eq:agr}, resulting in an asymmetry between the triplets in $\mathcal{D}^{V}$ and $\mathcal{D}$. Specifically, in $\mathcal{D}^{V}$, each user is associated with $n$ triplets, whereas in $\mathcal{D}$, each user has $1$ triplet. 
    \item $w/o$-Scale: This configuration removes the align-scale operation in Eq.~\ref{eq:loss} by modifying the loss function to $\mathcal{L} = \mathcal{L}_{bpr} + \lambda \mathcal{L}_{vbpr}$. This alteration causes the virtual triplet loss to introduce significant gradient skew when incorporated into the training process.
\end{itemize}

\begin{figure*}[!t]
    \centering 
    \includegraphics[width=1\linewidth]{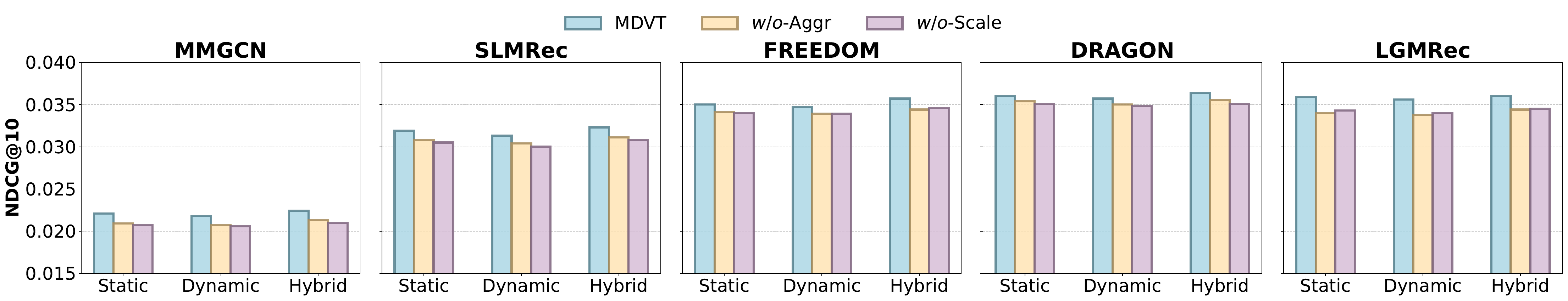}
     \vskip -0.15in
    \caption{Ablation study on key components of MDVT in terms of NDCG@10.}
    \label{fig:ab1}
     \vskip -0.15in
\end{figure*}

We conduct extensive experiments for our MDVT across five multimodal recommendation models on the Baby dataset for various configurations. The findings presented in Figure~\ref{fig:ab1} clearly demonstrate that our MDVT surpasses all its modified configurations, thereby confirming the essential role each component plays in learning high-quality representations. We believe that the inferior performance of all configurations compared to MDVT is due to the discrepancy in scale between the model loss during epochs when virtual triplets participate in joint optimization and the training loss during the warm-up epochs. This discrepancy leads to gradient skew, which is similar to that observed in multi-task learning \cite{li2018visualizing,yu2020gradient}. To validate our statement, we further investigate the changes in model loss and recommendation performance throughout the training phase. Specifically, we visualize the training loss and recommendation performance (NDCG@10) for two multimodal recommendation models with our MDVT and its configurations with the static warm-up threshold strategy (20 epoch warm-up) on the Baby dataset. The experimental results presented in Figure 3 support our statement. Specifically, compared with MDVT, all configurations exhibit a significant increase in loss and a noticeable decrease in performance after 20 warm-up epochs. Furthermore, their subsequent training is more unstable and requires longer training times than MDVT. These observations indicate that our key component design allows virtual triplets to be added to the training process without causing significant gradient skew, thereby maintaining stability and efficiency. 

Furthermore, we verify the effect of virtual triples construction. Specifically, we conducted an ablation study with various configurations: a) MDVT$_{ID}$, which builds virtual triples via only ID modality. b) MDVT$_{V}$, which builds virtual triples via only visual modality. c) MDVT$_{T}$, which builds virtual triples via only textual modality. d) MDVT$_{ID-V}$, which builds virtual triples via ID and visual modalities. e) MDVT$_{ID-T}$, which builds virtual triples via ID and textual modalities. f) MDVT$_{V-T}$, which builds virtual triples via both visual and textual modalities. g) MDVT$_{F1}$, which builds virtual triplets directly based on high and low interaction frequencies. g) MDVT$_{F2}$, which top-$2n$ virtual triplets based on representation similarity, selects the top-$n$ triplets according to interaction frequency. All experiments adopt the hybrid strategy. According to experimental results in Table~\ref{tab:ablation}, we have the following observations. 1) Models using only visual/textual modalities perform even worse than the original model, as the lack of ID causes the generated virtual triplets to deviate from the recommendation task. 2) The variant using only ID outperforms the original model, demonstrating the dominant role of ID. 3) The performance is further improved when ID is combined with visual/textual modalities. This suggests that the auxiliary modalities provide more modality information. 4) Constructing virtual triplets based on interaction frequency reduces the personalization of recommendations, thereby degrading the overall recommendation performance. 5) MDVT achieves the best performance, attributed to its ability to fuse informative information from multiple modalities to achieve optimal performance.

\begin{table*}[!ht]
\caption{Performance comparison for variants on three datasets in terms of Recall@10 (R@10).}
 \small
  \vskip -0.14in
\centering
\label{tab:ablation}
    \begin{tabular}{cc|cccccccccc}
    \toprule
        Dataset & Model & Original & MDVT$_{ID}$ &  MDVT$_{V}$ &  MDVT$_{T}$ &  MDVT$_{ID-V}$ & MDVT$_{ID-T}$ &  MDVT$_{V-T}$ & MDVT$_{F1}$& MDVT$_{F2}$ & MDVT \\ \midrule
        \multirow{5}{*}{Baby} & MMGCN & 0.0378 & 0.0381 & 0.0360 & 0.0370 & 0.0401 & 0.0412 & 0.0367& 0.0354 & 0.0410 & \textbf{0.0420} \\ 
        & SLMRec & 0.0529 & 0.0533 & 0.0508 & 0.0519 & 0.0545 & 0.0552 & 0.0515 & 0.0493 & 0.0550 & \textbf{0.0563} \\ 
        & FREEDOM & 0.0627 & 0.0631 & 0.0603 & 0.0617 & 0.0641 & 0.0649 & 0.0613 & 0.0589 & 0.0654 & \textbf{0.0662} \\ 
        & DRAGON & 0.0662 & 0.0666 & 0.0644 & 0.0653 & 0.0675 & 0.0682 & 0.0649 & 0.0627 & 0.0688 & \textbf{0.0692} \\ 
        & LGMRec & 0.0626 & 0.0633 & 0.0602 & 0.0615 & 0.0643 & 0.0652 & 0.0611 & 0.0522 & 0.0652 & \textbf{0.0660} \\ \midrule
        \multirow{5}{*}{Sports} & MMGCN & 0.0370 & 0.0375 & 0.0355 & 0.0362 & 0.0393 & 0.0401 & 0.0359 & 0.0341 & 0.0399 & \textbf{0.0406} \\ 
        & SLMRec & 0.0663 & 0.0668 & 0.0645 & 0.0653 & 0.0690 & 0.0702 & 0.0651 & 0.0619 & 0.0699 & \textbf{0.0709} \\ 
        & FREEDOM & 0.0717 & 0.0723 & 0.0696 & 0.0707 & 0.0741 & 0.0751 & 0.0703 & 0.0678 & 0.0751 & \textbf{0.0757} \\ 
        & DRAGON & 0.0752 & 0.0757 & 0.0736 & 0.0746 & 0.0770 & 0.0778 & 0.0744 & 0.0699 & 0.0773 & \textbf{0.0788} \\ 
        & LGMRec & 0.0719 & 0.0724 & 0.0700 & 0.0711 & 0.0749 & 0.0762 & 0.0708 & 0.0675 & 0.0762 & \textbf{0.0771} \\ \midrule
        \multirow{5}{*}{Clothing} & MMGCN & 0.0218 & 0.0225 & 0.0202 & 0.0211 & 0.0232 & 0.0237 & 0.0208 & 0.0200 & 0.0239 & \textbf{0.0244} \\
        & SLMRec & 0.0452 & 0.0461 & 0.0438 & 0.0446 & 0.0481 & 0.0490 & 0.0445 & 0.0420 & 0.0485 & \textbf{0.0497} \\
        & FREEDOM & 0.0629 & 0.0636 & 0.0615 & 0.0623 & 0.0649 & 0.0657 & 0.0621 & 0.0580 & 0.0655 & \textbf{0.0665} \\ 
        & DRAGON & 0.0671 & 0.0681 & 0.0653 & 0.0663 & 0.0692 & 0.0701 & 0.0660 & 0.0633 & 0.0704 & \textbf{0.0713} \\ 
        & LGMRec & 0.0555 & 0.0559 & 0.0538 & 0.0549 & 0.0596 & 0.0610 & 0.0545 & 0.0518 & 0.0614 & \textbf{0.0622} \\ \bottomrule
    \end{tabular}
      \vskip -0.15in
\end{table*}

\begin{figure}[!t]
\centering
    \subfigure[FREEDOM-MDVT] {
        \label{fig:lp_mdvt_freedom}
        \includegraphics[width=0.32\linewidth]{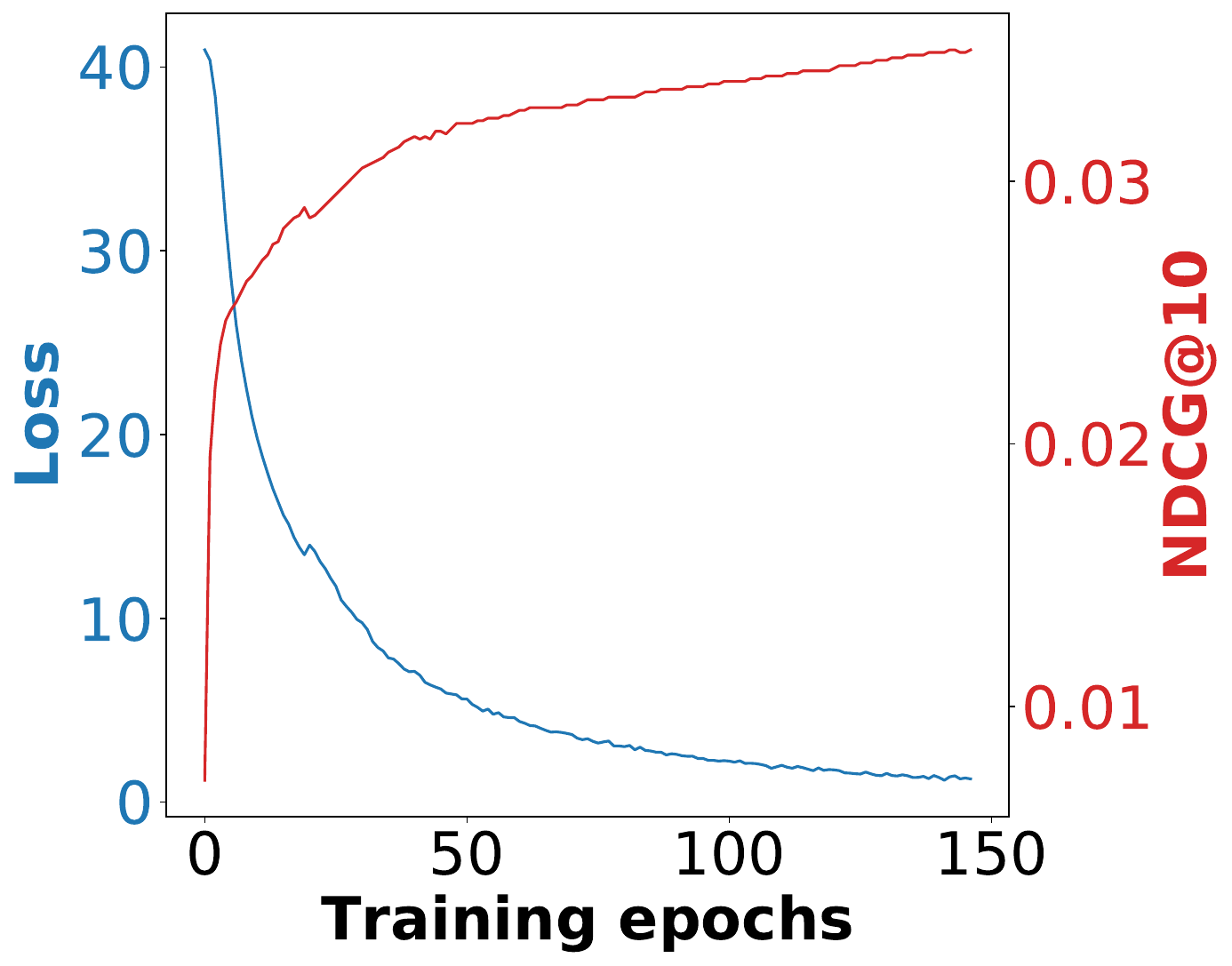}
        }  \hspace{-4mm}
    \subfigure[FREEDOM-$w/o$-Aggr] {
        \label{fig:lp_mdvtA_freedom}
        \includegraphics[width=0.32\linewidth]{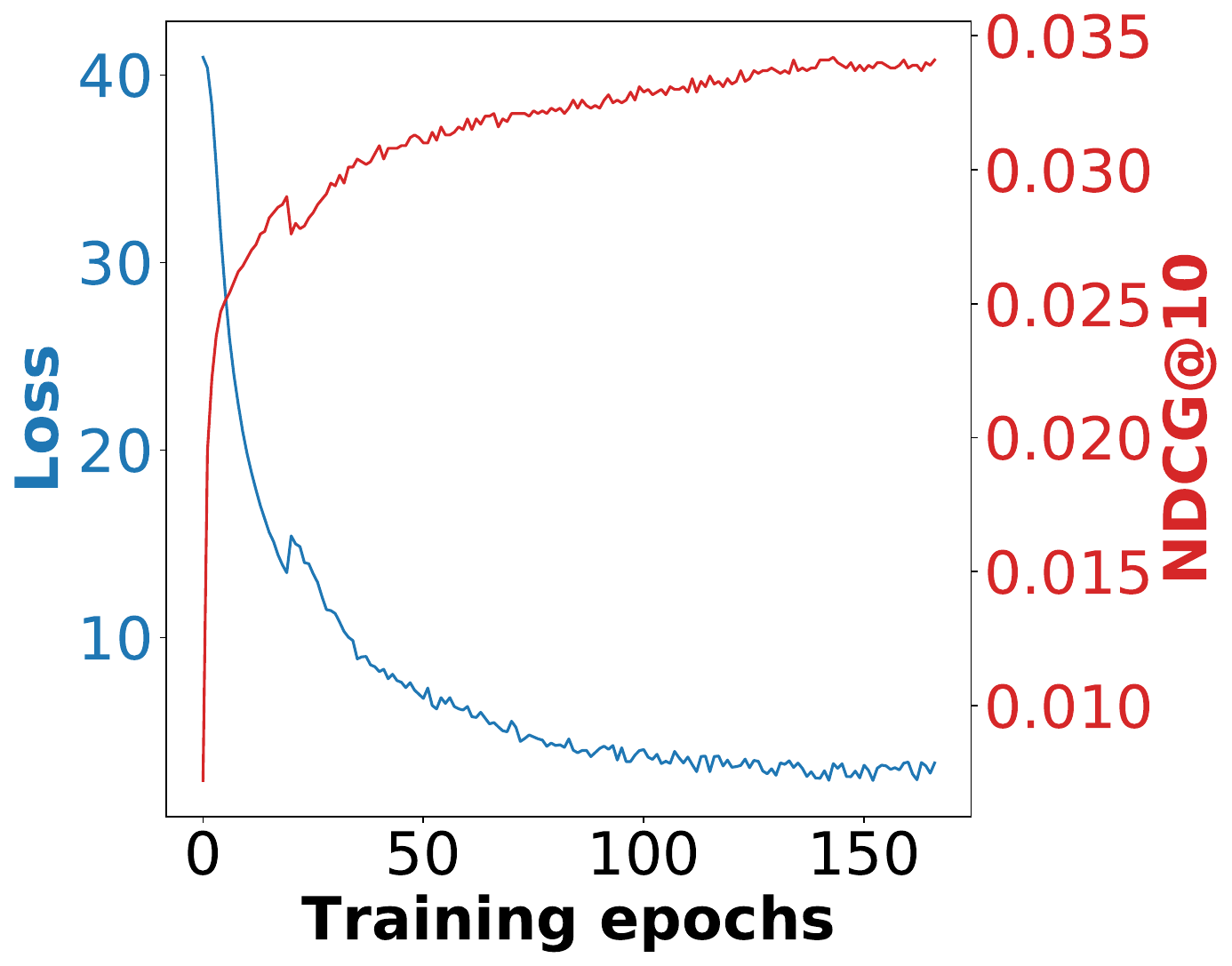}
        }  \hspace{-4mm}
    \subfigure[FREEDOM-$w/o$-Scale] {
        \label{fig:lp_mdvtS_freedom}
        \includegraphics[width=0.32\linewidth]{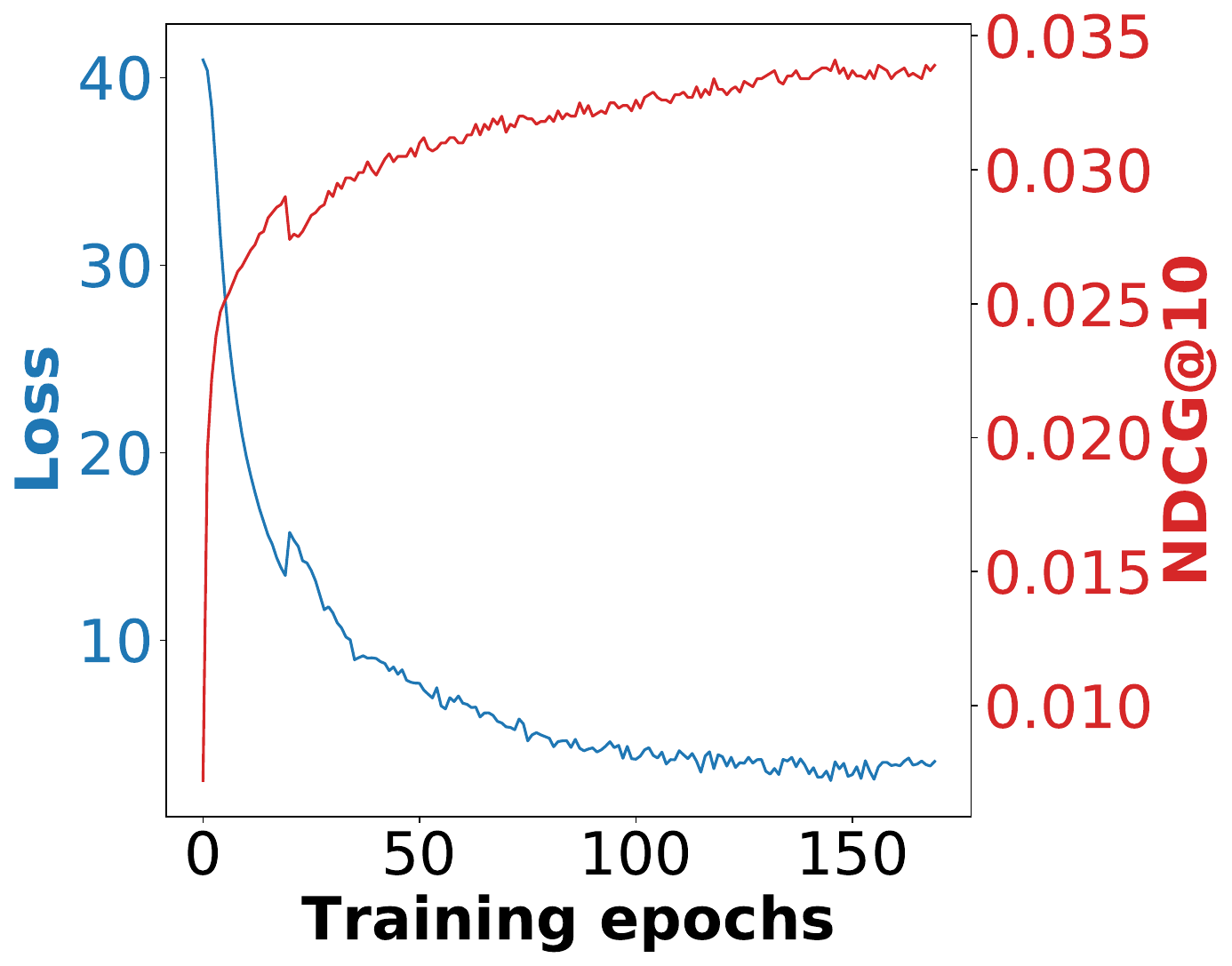}
        } \vskip -0.1in
    \subfigure[DRAGON-MDVT] {
        \label{fig:lp_mdvt_dragon}
        \includegraphics[width=0.32\linewidth]{Fig/lp_mdvt_freedom.pdf}
        }  \hspace{-4mm}
    \subfigure[DRAGON-$w/o$-Aggr] {
        \label{fig:lp_mdvtA_dragon}
        \includegraphics[width=0.32\linewidth]{Fig/lp_mdvtA_freedom.pdf}
        }  \hspace{-4mm}
    \subfigure[DRAGON-$w/o$-Scale] {
        \label{fig:lp_mdvtS_dragon}
        \includegraphics[width=0.32\linewidth]{Fig/lp_mdvtS_freedom.pdf}
        }    
    \vskip -0.2in
    \caption{The learning curve when adopting MDVT and its configurations to optimize the loss on the Baby dataset.}   
    \label{fig:l-p}   
    \vskip -0.2in
\end{figure}


\subsection{Sparsity Study (RQ3)}
\begin{figure}[!t]
    \centering
    \includegraphics[width=1\linewidth]{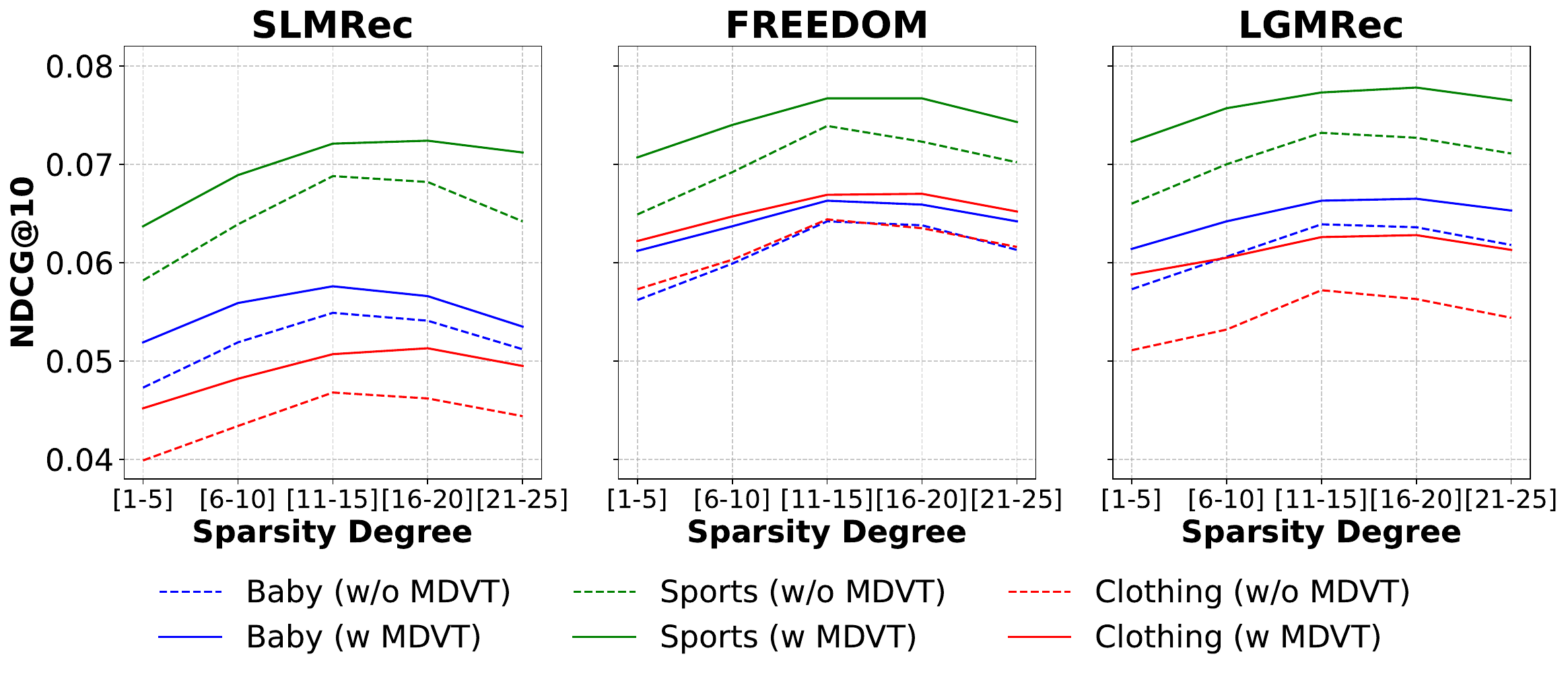}
     \vskip -0.17in
    \caption{Sparsity study on three advanced multimodal recommendation models across three distinct datasets.}
     \vskip -0.2in
    \label{fig:RQ3}
\end{figure}

To evaluate the effectiveness of adopting MDVT in advanced multimodal recommendation models under various data sparsity scenarios, we conduct experiments on subsets of all three datasets with differing sparsity levels. In particular, we compare the performance of three advanced multimodal recommendation models—SLMRec, FREEDOM, and LGMRec—with and without our MDVT. To analyze the effect of data sparsity, we categorize user groups based on their interaction counts in the training set (e.g., the first group consists of users who have interacted with 1-5 items). As illustrated in Figure~\ref{fig:RQ3}, MDVT consistently enhances the performance of these models across all datasets and sparsity levels, thereby confirming its effectiveness in diverse sparse scenarios. Furthermore, the improvement in recommendation performance of all models with MDVT is particularly significant in sparse scenarios, specifically for users with 1-5 and 6-10 interacted items. We attribute this enhancement to our virtual triplets being especially effective in sparse scenarios. 

\subsection{Convergence Speed (RQ4)}
\begin{figure}[!t]
    \centering
    \includegraphics[width=1\linewidth]{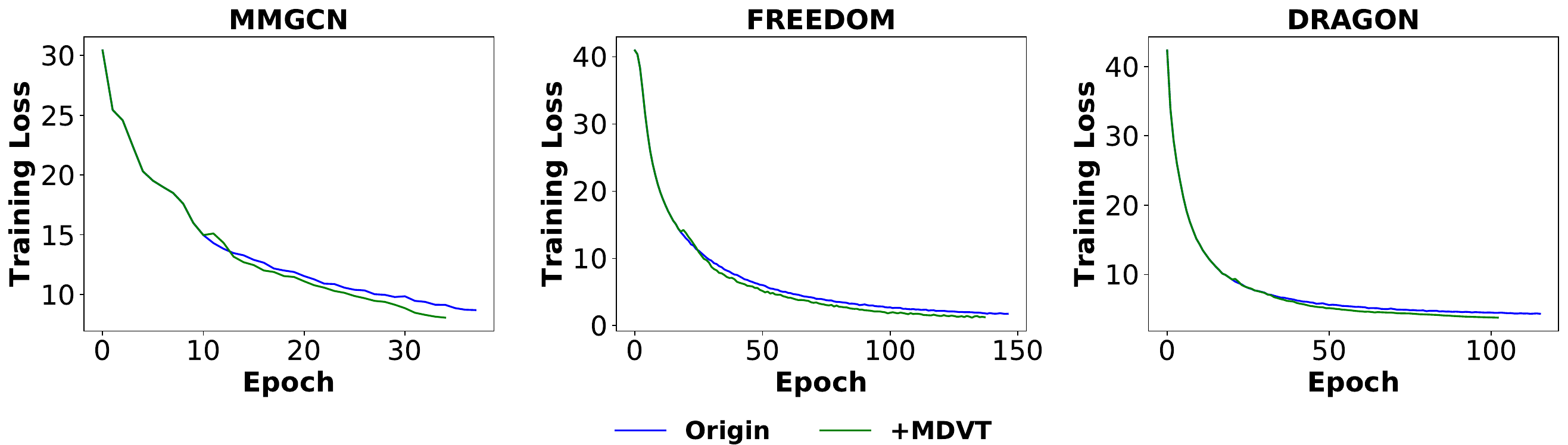}
    \vskip -0.15in
    \caption{Convergence study on the Baby dataset.}
    \vskip -0.18in
    \label{fig:RQ4}
\end{figure}

In addition, MDVT helps accelerate model training convergence. We visualized the training loss of three advanced multimodal recommendation models (MMGCN, FREEDOM, and DRAGON) on the Baby dataset. Following previous training settings \cite{zhou2023bootstrap,zhou2023tale}, we used an early stopping strategy with the patience of 20 epochs and set the maximum number of epochs to 1,000. As shown in Figure~\ref{fig:RQ4}, MDVT effectively improves the convergence speed of all models. We attribute this improvement to the virtual triplets providing informative supervision signals that accelerate model training convergence.

\subsection{Compatibility with Robust Training and Data Augmentation Strategies (RQ5)}
Existing studies enhance the robustness of multimodal recommendations by adversarial training strategy \cite{li2020adversarial,tang2019adversarial} and data augmentation method \cite{luo2024integrating,huang2024large,wei2024llmrec}. Therefore, we further evaluate the compatibility of our MDVT with the adversarial training strategy (AMR \cite{tang2019adversarial}) and LLM-based data augmentation strategy (GPT-4o \cite{yang2023dawn}). We conducted extensive experiments based on two multimodal recommendation models across three public datasets. Table~\ref{tab:compatible} shows that combining MDVT with both AMR and GPT-4o can further improve model performance. Additionally, GPT-4o outperforms AMR on all datasets, which we attribute to GPT-4o's ability to reduce the inherent gap between visual and textual information of items. Notably, simultaneously using both AMR and GPT-4o achieves more satisfactory performance than adopting either one alone.

\begin{table}[!t]
\caption{Performance comparison for strategies on three datasets under Recall@5 (R@5) and NDCG@5 (N@5). +M, +A, and +G denote +MDVT, +AMR, and +GPT-4o, respectively.}
  \vskip -0.15in
\centering
\tabcolsep=0.04in
\label{tab:compatible}
\resizebox{\linewidth}{!}{
    \begin{tabular}{c|c|cc|cc|cc}
    \toprule
         \multirow{2.5}{*}{Models} & Datasets& \multicolumn{2}{c|}{Baby}& \multicolumn{2}{c|}{Sports} & \multicolumn{2}{c}{Clothing}\\\cmidrule{2-8}
         & Metrics& R@5& N@5& R@5& N@5& R@5& N@5\\\midrule
         \multirow{5}{*}{MMGCN} & origin & 0.0240& 0.0160& 0.0216& 0.0143& 0.0130& 0.0110\\
         & +M & 0.0257& 0.0170& 0.0236& 0.0158& 0.0148& 0.0099\\
         & +M+A & 0.0260& 0.0172& 0.0239& 0.0160& 0.0150& 0.0101\\ 
         & +M+G & \underline{0.0266}& \underline{0.0176}& \underline{0.0244}& \underline{0.0163}& \underline{0.0155}& \underline{0.0104}\\ 
         & +M+A+G & \textbf{0.0268}& \textbf{0.0177}& \textbf{0.0246}& \textbf{0.0164}& \textbf{0.0157}& \textbf{0.0105}\\ \midrule
         \multirow{5}{*}{LGMRec} & origin & 0.0374& 0.0249& 0.0446& 0.0288& 0.0371& 0.0246\\
         & +M & 0.0417& 0.0281& 0.0475& 0.0312& 0.0411& 0.0276\\
         & +M+A & 0.0420& 0.0282& 0.0477& 0.0313& 0.0414& 0.0278\\ 
         & +M+G & \underline{0.0427}& \underline{0.0288}& \underline{0.0486}& \underline{0.0318}& \underline{0.0419}& \underline{0.0281}\\ 
         & +M+A+G & \textbf{0.0431}& \textbf{0.0291}& \textbf{0.0488}& \textbf{0.0320}& \textbf{0.0421}& \textbf{0.0283}\\ \bottomrule
    \end{tabular}
    }
    \vskip -0.15in
\end{table}

\subsection{Mechanism for Threshold Strategies (RQ6)}
\label{sec:mech}
\begin{figure*}
    \centering
    \includegraphics[width=1\linewidth]{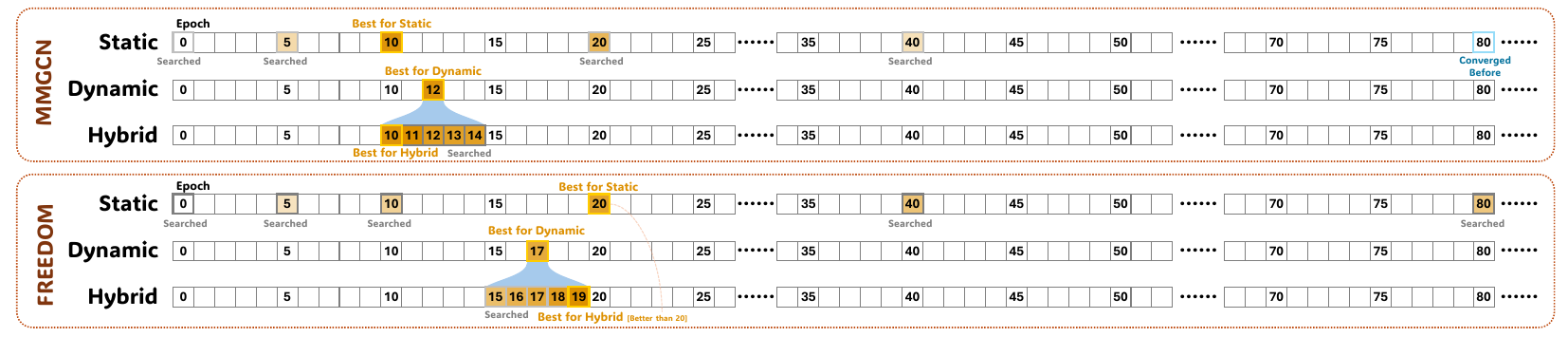}
    \vskip -0.15in
    \caption{Mechanics of all three warm-up threshold strategies for two advanced models on the Baby dataset.}
     \vskip -0.15in
    \label{fig:mech}
\end{figure*}

We further explore the practical training with these three warm-up threshold strategies. We conduct the hyper-parameter search for two advanced models (MMGCN and FREEDOM) with these three strategies on the Baby dataset. For the static warm-up threshold strategy, we follow the search range introduced in Section~\ref{sec:impl}. For the dynamic warm-up threshold strategy, we set $g$ = 0.1 (as shown in Section~\ref{sec:hyper}, $g$ = 0.1 or 0.2 can be applied to all datasets). For the hybrid warm-up threshold strategy, we first applied the dynamic strategy with $g$ = 0.1 to estimate the approximately optimal warm-up epochs $\mathcal{T}^{cur}$. Then we adopt the static strategy within the range $[\mathcal{T}^{cur}-s, \mathcal{T}^{cur}+s]$, where $s$ = 2. As shown in Figure~\ref{fig:mech}, the optimal warm-up epochs for all three strategies are within a similar range. Moreover, the hybrid strategy combines the advantages of both static and dynamic strategies, achieving satisfactory performance with available hyper-parameter adjustment overhead.

\subsection{Hyper-parameter Analysis (RQ7)}
\label{sec:hyper}
We evaluate the impact of the key hyper-parameters ($\lambda$, $n$, $g$, and $s$) on MDVT's performance across three Amazon datasets in terms of Recall@10. For the hyper-parameters $\lambda$ and $n$, we conduct analyses based on the hybrid warm-up threshold strategy, as these hyper-parameters are not related to the choice of warm-up threshold strategy, and the hybrid warm-up threshold strategy has demonstrated superior performance over the static and dynamic warm-up threshold strategies. For the hyper-parameter $g$, which is contained in both the dynamic and hybrid warm-up threshold strategies, we provide analyses based on these two strategies. Similarly, for the hyper-parameter $s$, we provide analysis based on the hybrid warm-up threshold strategy, which is the only strategy that contains $s$. 

\begin{figure}[!t]
    \centering
    \includegraphics[width=1\linewidth]{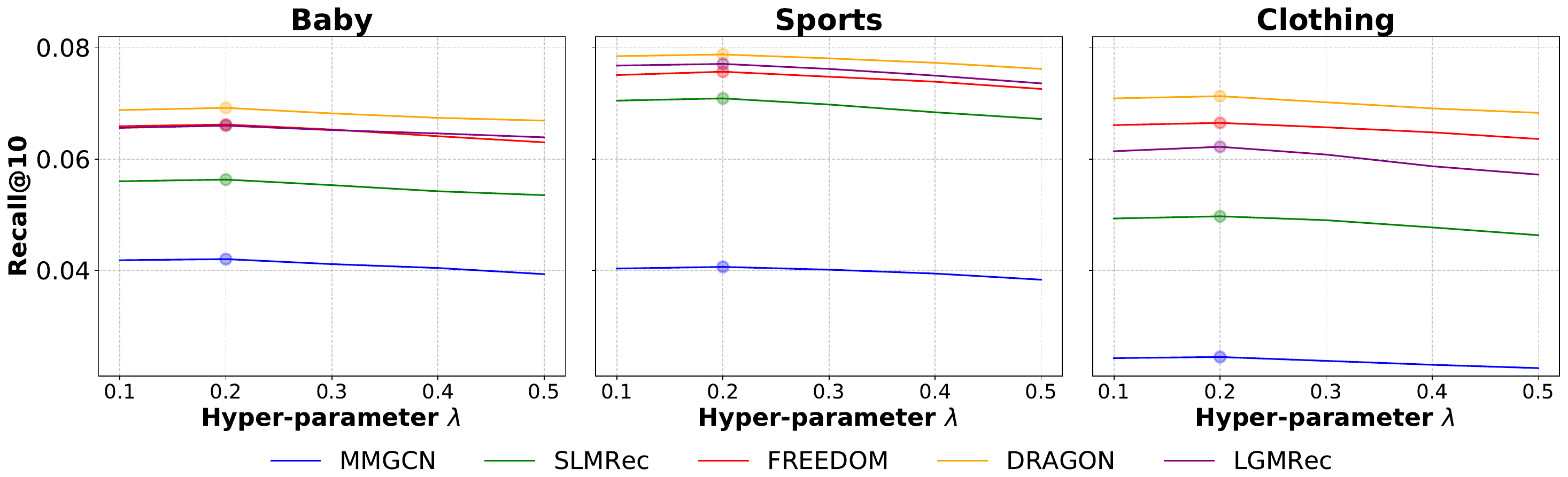}
    \vskip -0.15in
    \caption{Performance $w.r.t.$ hyper-parameter $\lambda$.}
    \vskip -0.18in
    \label{fig:RQ7_1}
\end{figure}
\begin{figure}[!t]
    \centering
    \includegraphics[width=1\linewidth]{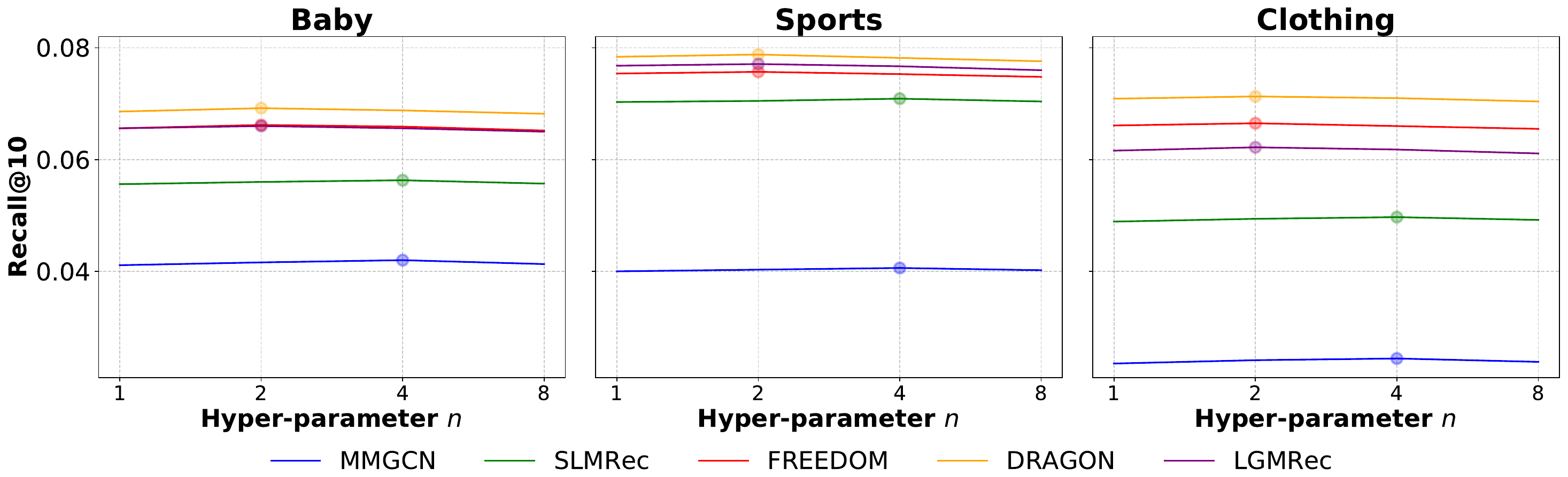}
    \vskip -0.15in
    \caption{Performance $w.r.t.$ hyper-parameter $n$.}
    \vskip -0.18in
    \label{fig:RQ7_2}
\end{figure}

\noindent\textbf{Hyper-parameter $\lambda$ and $n$}: From Figure~\ref{fig:RQ7_1} and Figure~\ref{fig:RQ7_2}, we have the following observations. For FREEDOM, DRAGON, and LGMRec across all datasets, the optimal hyper-parameters $\lambda$ and $n$ are 0.2 and 2, respectively. In contrast, for MMGCN and SLMRec, the optimal hyper-parameters are higher, with $\lambda$ = 0.2 and $n$ = 4 across all datasets. We attribute this phenomenon to the lower baseline performance of MMGCN and SLMRec compared to the other models. These models are more affected by the data sparsity problem and thus require a larger value of $n$ to fully leverage virtual triplets for performance enhancement.

\begin{figure}[!t]
    \centering
    \includegraphics[width=1\linewidth]{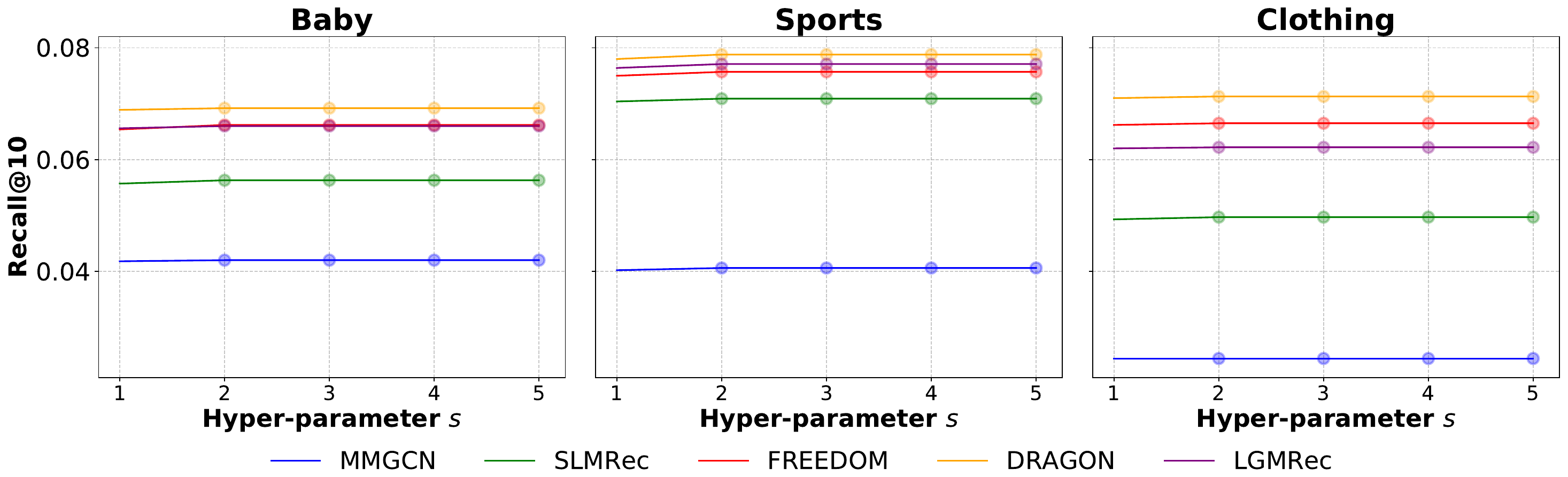}
    \vskip -0.15in
    \caption{Performance $w.r.t.$ hyper-parameter $s$.}
    \vskip -0.18in
    \label{fig:RQ7_4}
\end{figure}

\noindent\textbf{Hyper-parameter $g$ and $s$}: From Table~\ref{tab:hyper} and Figure~\ref{fig:RQ7_4}, we have the following observations. For the hyper-parameter $g$, values of 0.1 and 0.2 are recommended for all five advanced multimodal recommendation models across all datasets. For the hyper-parameter $s$, a value of 2 is sufficient to find the optimal number of warm-up epochs for all models. Therefore, we conclude that the hybrid warm-up threshold strategy can achieve satisfactory enhancements without incurring high hyper-parameter tuning costs.

\begin{table}[!t]
\caption{Analysis for hyper-parameter $g$ for dynamic and hybrid strategies based on four multimodal recommendation models across all datasets in terms of Recall@10.}
  \vskip -0.1in
\centering
\tabcolsep=0.04in
\label{tab:hyper}
\resizebox{\linewidth}{!}{
    \begin{tabular}{c|c|cc|cc|cc}
    \toprule
         \multirow{2}{*}{Models} & Datasets& \multicolumn{2}{c|}{Baby}& \multicolumn{2}{c|}{Sports} & \multicolumn{2}{c}{Clothing}\\\cmidrule{2-8}
         & Strategies& Dynamic& Hybrid& Dynamic& Hybrid& Dynamic& Hybrid\\\midrule
         \multirow{5}{*}{MMGCN} & $g$ = 0.1& \underline{0.0411}& \underline{0.0416}& \underline{0.0397}& \underline{0.0404}& \underline{0.0240}& \underline{0.0242}\\
         & $g$ = 0.2& \textbf{0.0413}& \textbf{0.0420}& \textbf{0.0400}& \textbf{0.0406}& \textbf{0.0244}& \textbf{0.0244}\\
         & $g$ = 0.3& 0.0401& 0.0404& 0.0390& 0.0393& 0.0231& 0.0236\\
         & $g$ = 0.4& 0.0382& 0.0388& 0.0373& 0.0379& 0.0220& 0.0225\\
         & $g$ = 0.5& 0.0365& 0.0368& 0.0359& 0.0362& 0.0213& 0.0217\\ \midrule
         \multirow{5}{*}{FREEDOM} & $g$ = 0.1& \textbf{0.0648}& \textbf{0.0662}& \textbf{0.0750}& \textbf{0.0757}& \underline{0.0653}& \underline{0.0663}\\
         & $g$ = 0.2& \underline{0.0643}& \underline{0.0655}& \underline{0.0747}& \underline{0.0752}& \textbf{0.0655}& \textbf{0.0655}\\
         & $g$ = 0.3& 0.0635& 0.0643& 0.0736& 0.0743& 0.0641& 0.0647\\
         & $g$ = 0.4& 0.0622& 0.0628& 0.0720& 0.0726& 0.0626& 0.0630\\
         & $g$ = 0.5& 0.0610& 0.0618& 0.0703& 0.0708& 0.0618& 0.0620\\ \midrule
         \multirow{5}{*}{DRAGON} & $g$ = 0.1& \underline{0.0680}& \underline{0.0688}& \textbf{0.0776}& \textbf{0.0788}& \underline{0.0701}& \underline{0.0708}\\
         & $g$ = 0.2& \textbf{0.0685}& \textbf{0.0692}& \underline{0.0774}& \underline{0.0785}& \textbf{0.0704}& \textbf{0.0713}\\
         & $g$ = 0.3& 0.0674& 0.0679& 0.0765& 0.0773& 0.0692& 0.0699\\
         & $g$ = 0.4& 0.0663& 0.0669& 0.0757& 0.0769& 0.0678& 0.0678\\
         & $g$ = 0.5& 0.0648& 0.0655& 0.0742& 0.0750& 0.0653& 0.0661\\ \midrule
         \multirow{5}{*}{LGMRec} & $g$ = 0.1& \underline{0.0648}& \underline{0.0655}& \underline{0.0758}& \underline{0.0767}& \textbf{0.0612}& \textbf{0.0622}\\
         & $g$ = 0.2& \textbf{0.0651}& \textbf{0.0660}& \textbf{0.0762}& \textbf{0.0771}& \underline{0.0610}& \underline{0.0618}\\
         & $g$ = 0.3& 0.0638& 0.0643& 0.0745& 0.0753& 0.0593& 0.0599\\
         & $g$ = 0.4& 0.0625& 0.0627& 0.0724& 0.0728& 0.0570& 0.0577\\
         & $g$ = 0.5& 0.0618& 0.0621& 0.0702& 0.0708& 0.0548& 0.0553\\ \midrule
    \end{tabular}
    }
    \vskip -0.2in
\end{table}

\section{Related Work and Model Details}
Due to page limitations, we review recent works and their contributions in Appendix~\ref{appendix:related work}. Moreover, we provide details for all utilized models in Appendix~\ref{appendix:models}. More discussions are in Appendix~\ref{appendix:discussion}.

\section{Conclusion}
In this paper, we aimed to mitigate the data sparsity problem in multimodal recommendation systems by leveraging multimodal information more effectively. We propose a novel, model-agnostic approach called MDVT, which constructs \textbf{M}ultimodal-\textbf{D}riven \textbf{V}irtual \textbf{T}riplets to provide valuable supervision signals for model training. To ensure the high quality of these virtual triplets, we introduce three different warm-up threshold strategies tailored to fit various real-world scenarios. Once the warm-up threshold is satisfied, the virtual triplets are used for joint model optimization, enhancing the learning process without causing significant gradient skew. MDVT is model-agnostic and can be easily integrated into any multimodal recommendation model. Extensive experiments on multiple real-world datasets across various advanced models demonstrated the effectiveness of MDVT. These results confirm that leveraging virtual triplets can significantly improve recommendation performance by alleviating the data sparsity problem. In future work, we aim to develop representation enhancement techniques to improve the quality of virtual triplets, thereby enhancing supervision signals and boosting overall model performance.


\begin{acks}
This work was supported by the Hong Kong UGC General Research Fund no. 17203320 and 17209822, and the project grants from the HKU-SCF FinTech Academy.
\end{acks}

\balance

\bibliographystyle{ACM-Reference-Format}


\newpage

\appendix

\section{Appendix}

\subsection{Algorithm and Validation}
\label{appendix:algorithm}
We present the procedure in Algorithm~\ref{alg:all}, which integrates our model-agnostic MDVT into the multimodal recommendation model with three different warm-up threshold strategies. Note that hyper-parameters $g$ and $s$ can be easily defined by 0.1 and 2, respectively, to achieve ideal performance for all models across all datasets, which are validated empirically in Section~\ref{sec:hyper} and Section~\ref{sec:mech}. 

We validate the effectiveness of these three threshold strategies in Section~\ref{subsec:RQ1}. Furthermore, we provide an in-depth analysis of the mechanism for these three threshold strategies in Section~\ref{sec:mech}. Our code link can be found in the footnote 3.
\begin{algorithm}
\caption{Procedure with Threshold Strategies}
\label{alg:all}
\KwIn{Strategy Type $type$, Threshold Set $\mathcal{S_T}$, Pre-defined Hyper-parameter $g$ and $s$;}
\KwOut{Optimal warm-up epochs $\mathcal{T_O}$;}
Auxiliary Markings: $\mathcal{P}^{bar} = 0$, $\mathcal{T}^{cur}$, and $flag = \mathbf{False}$;\\
\While{$flag = \mathbf{True}$}{
Initialize model parameters $\Theta$ and $\mathcal{L}^{p} = 0$;\\
\If{$type$ $=$ $Dynamic$ or $Hybrid$}{
$\mathcal{T} = 0$, $f = \mathbf{False}$;\\
\While{not converged}{
$\mathcal{T} = \mathcal{T}+1$;\\
Get fused representations via Eq~\ref{eq:1},\ref{eq:2}, and \ref{eq:4};\\
Calculate BPR loss $\mathcal{L}_\mathrm{bpr}$ via Eq~\ref{eq:bpr}; \\
\If{$\mathcal{L}^{p} \neq 0$, $f = \mathbf{False}$, and $\frac{\mathcal{L}^{bpr}-\mathcal{L}^{p}}{\mathcal{L}^{p}} \leq g$}{
    Update $\mathcal{T}^{cur} = \mathcal{T}$, $f = \mathbf{True}$;\\
}
\If{$f = \mathbf{True}$}{
Get virtual triplet dataset $\mathcal{D}^V$ via Eq~\ref{eq:5}-\ref{eq:6};\\
Calculate virtual loss $\mathcal{L}_\mathrm{vbpr}$ via Eq~\ref{eq:vbpr}-\ref{eq:agr};
}
Calculate final loss $\mathcal{L}$ with Eq~\ref{eq:loss}; \\
Update $\mathcal{L}^{prev}$ by current loss $\mathcal{L}$;\\
Calculate the gradient of loss $\nabla_{\Theta} \mathcal{L}$; \\
Update $\Theta$ by gradient $\nabla_{\Theta} \mathcal{L}$ with optimizer;
}
Test model performance $\mathcal{P}$;\\
Update $\mathcal{P}^{bar} = \mathcal{P}$, $\mathcal{S_T} = [\mathcal{T}^{cur}-s, \mathcal{T}^{cur}+s]$;\\
}
\If{$type$ $=$ $Static$ or $Hybrid$}{
\For{$\mathcal{T_S} \in \mathcal{S_T}$}{
    $\mathcal{T} = 0$;\\
\While{not converged}{
$\mathcal{T} = \mathcal{T}+1$;\\
Get fused representations via Eq~\ref{eq:1},\ref{eq:2}, and \ref{eq:4};\\
Calculate BPR loss $\mathcal{L}_\mathrm{bpr}$ via Eq~\ref{eq:bpr}; \\
\If{$\mathcal{T} >= \mathcal{T_S}$}{
    Get virtual triplet dataset $\mathcal{D}^V$ via Eq~\ref{eq:5}-\ref{eq:6};\\
    Calculate virtual loss $\mathcal{L}_\mathrm{vbpr}$ via Eq~\ref{eq:vbpr}-\ref{eq:agr};
}
Calculate final loss $\mathcal{L}$ with Eq~\ref{eq:loss}; \\
Update $\mathcal{L}^{prev}$ by current loss $\mathcal{L}$;\\
Calculate the gradient of loss $\nabla_{\Theta} \mathcal{L}$; \\
Update $\Theta$ by gradient $\nabla_{\Theta} \mathcal{L}$ with optimizer;
}
Test model performance $\mathcal{P}$;\\
\If{$\mathcal{P} > \mathcal{P}^{bar}$}{
Update $\mathcal{P}^{bar} = \mathcal{P}$, $\mathcal{T}^{cur} = \mathcal{T_S}$;\\
}
}
}
Update $\mathcal{T_O} = \mathcal{T}^{cur}$;
}

\end{algorithm}

\subsection{Related Work}
\label{appendix:related work}
Recent studies incorporate multimodal information to mitigate the data sparsity problem in recommendation systems. Pioneering this approach, VBPR \cite{he2016vbpr} leverages visual content as side information in matrix factorization \cite{rendle2009bpr}, utilizing item images to enhance recommendations. Building upon this foundation, subsequent works \cite{chen2019personalized,liu2019user,yu2023multi,chen2025don} further integrate both visual and textual modalities to enrich item representations and improve performance. Advancements in graph-based methods introduce new avenues for multimodal recommendations. MMGCN \cite{wei2019mmgcn} is the first to integrate Graph Convolutional Networks (GCNs) to extract modality-specific features from user-item interactions. To explicitly capture commonalities in user preferences and item relationships, models like DualGNN \cite{wang2021dualgnn} and LATTICE \cite{zhang2021mining} leverage user-user and item-item graphs, respectively. Building on LATTICE, FREEDOM \cite{zhou2023tale} further stabilizes representations by freezing the item semantic graph and reducing noise in the user-item bipartite graph. Recently, self-supervised learning and inter-modal relationships have been explored to enhance recommendation systems. MMSSL \cite{wei2023multi} and MENTOR \cite{xu2025mentor} employ contrastive self-supervised learning to align modalities with collaborative signals, improving performance without extensive labelled data. Additionally, BM3 \cite{zhou2023bootstrap} investigates inter-modal relationships to boost recommendation accuracy and modality fusion quality. Furthermore, LGMRec \cite{guo2024lgmrec} leverages hyper-graph to capture complex global and local relationships in multimodal information. COHESION \cite{xu2025cohesion} design a tailored dual-stage fusion to boost multimodal recommendation performance.


While existing works typically employ multimodal information only as side information to model user preferences, we propose leveraging the similarity between user and item modality representations, which can also provide valuable supervision signals beyond explicit user-item interactions. We construct virtual triplets based on multimodal information to capitalize on this, providing informative supervision signals to mitigate the data sparsity problem.

\subsection{Models}
\label{appendix:models}
In this section, we introduce the five advanced multimodal models used for evaluation:
\begin{itemize}[leftmargin=*]
\item \textbf{MMGCN} \cite{wei2019mmgcn} applies GCN for each data modality to learn modality-specific features and then integrates all user-predicted ratings across modalities to produce the final rating.
\item \textbf{SLMRec} \cite{tao2022self} leverages a self-supervised learning framework for multimodal recommendations by establishing a tailored node self-discrimination task, which reveals hidden multimodal patterns.
\item \textbf{FREEDOM} \cite{zhou2023tale} refines LATTICE by freezing the item-item graph to stabilize item relationships and reducing noise in the user-item graph to enhance recommendation accuracy.
\item \textbf{DRAGON} \cite{zhou2023enhancing} leverages heterogeneous and homogeneous graphs to learn high-quality user/item representations.
\item \textbf{LGMRec} \cite{guo2024lgmrec} integrates local embeddings, which capture fine-grained topological embeddings, with global embeddings considering hypergraph dependencies among items.
\item \textbf{MMSSL} \cite{wei2023multi} combines modality-aware adversarial training with cross-modal contrastive learning to learn both cross-modality and modality-specific features.
\end{itemize}

\subsection{More Discussion}
\label{appendix:discussion}

\begin{table*}[!ht]
    \caption{Performance comparison for variants on the Baby dataset in terms of NDCG@10.}
 \small
  \vskip -0.15in
  \label{tab:appendix1}
\centering
    \begin{tabular}{c|ccccc}
    \toprule
        Baby & MMGCN & SLMRec & FREEDOM & DRAGON & LGMRec \\ \midrule
        MDVT & \textbf{0.0224} & \textbf{0.0323} & \textbf{0.0357} & \textbf{0.0364} & \textbf{0.0360} \\ 
        MDVT $w$ p $w/o$ n & 0.0212 (0.6) & 0.0315 (0.6) & 0.0354 (0.6) & 0.0360 (0.6) & 0.0358 (0.6) \\ 
        MDVT $w$ p $w/o$ w,n & 0.0200 (0.9) & 0.0287 (0.9) & 0.0337 (0.7) & 0.0353 (0.7) & 0.0339 (0.7) \\ 
        Original & 0.0200 & 0.0290 & 0.0330 & 0.0345 & 0.0333 \\ \bottomrule
    \end{tabular}
     \vskip -0.15in
\end{table*}

\begin{table*}[!ht]
    \caption{Performance comparison for variants on the Baby dataset in terms of NDCG@10.}
 \small
  \vskip -0.15in
  \label{tab:appendix2}
\centering
    \begin{tabular}{c|ccccc}
    \toprule
    Baby & MMGCN & SLMRec & FREEDOM & DRAGON & LGMRec \\\midrule
        MDVT & 0.0224 & 0.0323 & \textbf{0.0357} & \textbf{0.0364} & \textbf{0.0360} \\ 
        MDVT ($w$ p) & 0.0218 (0.9, 4) & 0.0318 (0.9, 4) & 0.0355 (0.7, 2) & \textbf{0.0364} (0.7, 2) & 0.0358 (0.7, 2) \\ 
        MDVT+ ($w$ p) & \textbf{0.0227} (0.9, [1,4]) & \textbf{0.0329} (0.9, [1,4]) & \textbf{0.0357} (0.7, [2,2]) & \textbf{0.0364} (0.7, [2,2]) & \textbf{0.0360} (0.7, [2,2]) \\ 
        Original & 0.0200 & 0.0290 & 0.0330 & 0.0345 & 0.0333 \\ \bottomrule
        \end{tabular}
     \vskip -0.15in
\end{table*}

Inspired by CC-GCN \cite{xv2023commerce}, we introduced a predefined threshold $\mathcal{T}$ to filter virtual triplets, hypothesizing that $\mathcal{T}$ acts as a dynamic warm-up strategy. Specifically, when the model is undertrained, user-item similarities are low, and few virtual triplets are constructed. To test this, we designed two MDVT variants: MDVT ($w$ p $w/o$ n) with a warm-up phase and MDVT ($w$ p $w/o$ w,n) without it, both using $\mathcal{T}$ instead of the top-$n$ strategy. NDCG@10 results on the Baby dataset (search space $\mathcal{T} \in {0.5, 0.6, 0.7, 0.8, 0.9}$) are reported, with best-performing $\mathcal{T}$ values in parentheses. According to experimental results in Table~\ref{tab:appendix1}, we observed that the MDVT variant without a warm-up phase (MDVT ($w$ p $w/o$ w,n)) led to negative optimization effects on SLMRec and showed less improvement on other models compared to MDVT ($w$ p $w/o$ n). Additionally, this variant required a higher threshold $\mathcal{T}$ to mitigate these issues. To further investigate, we analyzed the changes in similarity between user and item representations during the optimization process. We identified the following reasons: during early training epochs, the BPR loss fails to fully establish a user-item representation space, leading to disordered similarities and repeated selection of incorrect high-similarity items as virtual triplets, disrupting representation learning. A higher $\mathcal{T}$ mitigates this issue by reducing the impact of such errors. Advanced models like FREEDOM, DRAGON, and LGMRec construct user-item representations more effectively within 5-10 epochs, achieving minor improvements even without a warm-up phase. However, the MDVT variant with the warm-up phase ($w$ p $w/o$ n) partially avoids these issues but still underperforms compared to MDVT. Further analysis showed that for users with sparse interactions (1-3 records), the warm-up phase left few or no items meeting the $\mathcal{T}$ threshold, worsening popularity bias. Lowering $\mathcal{T}$ addressed this for sparse users but generated excessive virtual triplets for dense users, negatively affecting performance. 
Furthermore, we considered combining the top-$n$ strategy with the predefined threshold $\mathcal{T}$ to avoid generating excessive virtual triplets for users with dense interactions. We designed the following two variants: MDVT ($w$ p) and MDVT+ ($w$ p). The former employs the top-$n$ strategy to limit the number of virtual triplets constructed for users with dense interactions. The latter extends this method by replacing top-$n$ with an interval [n1, n2], ensuring that all users with sparse interactions can generate at least n1 virtual triplets, while users with dense interactions generate no more than n2 virtual triplets. We report the NDCG@10 results on the Baby dataset. For n1, the hyperparameter search range is $\{0, 1, 2\}$, and for n or n2, the search range is $\{1, 2, 4, 8\}$. In the table below, the numbers in parentheses indicate the optimal hyper-parameters selected for ($\mathcal{T}$, n) or ($\mathcal{T}$, [n1, n2]). Based on the experimental results, we summarize the following observations: the performance of MDVT ($w$ p) is inferior to both MDVT+ ($w$ p) and MDVT, which validates our earlier finding that 'for most users with sparse interactions (1-3 interaction records), after the warm-up phase, there are either no items or only a few items that satisfy the similarity threshold $\mathcal{T}$, thereby exacerbating the popularity bias.' Moreover, MDVT+ ($w$ p) is equivalent to MDVT on advanced models such as FREEDOM, DRAGON, and LGMRec. Additionally, it slightly outperforms MDVT on MMGCN and SLMRec. MMGCN and SLMRec have relatively weaker modeling capabilities, introducing noise for users with sparse interactions. Combining the predefined threshold $\mathcal{T}$ and top-$n$ strategy mitigates this issue.

In conclusion, our findings are as follows: 1) The predefined threshold $\mathcal{T}$ alone is insufficient for satisfactory MDVT performance due to fundamental differences from CC-GCN. While CC-GCN uses content-based similarity to construct virtual samples, MDVT dynamically builds virtual triplets based on evolving user-item representation similarities during optimization. 2) Combining the top-$n$ strategy with $\mathcal{T}$ outperforms MDVT on weaker models and is equivalent for advanced models. However, it requires an additional top-$n$ interval to ensure sufficient virtual triplets for users with sparse interactions and demands careful tuning of $\mathcal{T}$.

\end{document}